\documentclass[journal]{IEEEtran}
\usepackage{cite}

\ifCLASSINFOpdf
  \usepackage[pdftex]{graphicx}
  \graphicspath{{./images/}}
\else
\fi
\usepackage[cmex10]{amsmath}
\ifCLASSOPTIONcompsoc
  \usepackage[caption=false,font=normalsize,labelfont=sf,textfont=sf]{subfig}
\else
  \usepackage[caption=false,font=footnotesize]{subfig}
\fi
\usepackage{url}

\usepackage{epsfig}
\usepackage{multirow}
\usepackage[table,xcdraw,dvipsnames]{xcolor}

\newcolumntype{L}[1]{>{\raggedright\let\newline\\\arraybackslash\hspace{0pt}}m{#1}}
\newcolumntype{C}[1]{>{\centering\let\newline\\\arraybackslash\hspace{0pt}}m{#1}}
\newcolumntype{R}[1]{>{\raggedleft\let\newline\\\arraybackslash\hspace{0pt}}m{#1}}

\begin{document}


\title{Towards the Emulation of the Cardiac Conduction System for Pacemaker Testing}

%

\author{Eugene~Yip, Sidharta~Andalam, Partha~S.~Roop, Avinash~Malik, Mark~Trew, Weiwei~Ai, and Nitish Patel  
\thanks{E. Yip was and S. Andalam, P. S. Roop, A. Malik, W. Ai, and N. Patel are with the Department of Electrical and Computer Engineering, 
		the University of Auckland, New Zealand. M. Trew is with the Auckland Bioengineering Institute, the University of Auckland, New Zealand.}
\thanks{E-mails: \{eyip002, wai484\}@aucklanduni.ac.nz and \{sid.andalam, p.roop, avinash.malik, nd.patel, m.trew\}@auckland.ac.nz}%
}

%
%

\markboth{}%
{Yip \MakeLowercase{\textit{et al.}}: Towards the Emulation of the Cardiac Conduction System for Pacemaker Testing}
%



\maketitle

\begin{abstract}
	The heart is a vital organ that relies on the orchestrated propagation of
	electrical stimuli to coordinate each heart beat. Abnormalities in the
	heart's electrical behaviour can be managed with a cardiac
	pacemaker. 
	Recently, the closed-loop testing of pacemakers with an emulation 
	(real-time simulation) of the heart has been proposed. An emulated
	heart would provide realistic reactions to the pacemaker as if it were
	a real heart. This enables developers to interrogate their pacemaker
	design without having to engage in costly or lengthy clinical
	trials. Many high-fidelity heart models have been developed, but are
	too computationally intensive to be simulated in real-time. Heart
	models, designed specifically for the closed-loop testing of
	pacemakers, are too abstract to be useful in the testing of physical
	pacemakers.
	
	In the context of pacemaker testing, this paper presents
	a more computationally efficient heart model that generates 
	realistic continuous-time electrical signals. The heart 
	model is composed of cardiac cells that are connected by 
	paths. Significant improvements were made to an existing 
	cardiac cell model to stabilise its activation behaviour 
	and to an existing path model to capture the behaviour 
	of continuous electrical propagation. We provide simulation 
	results that show our ability to faithfully model complex 
	re-entrant circuits (that cause arrhythmia) that existing 
	heart models can not.
\end{abstract}

\begin{IEEEkeywords}
	cardiac, electrophysiology, emulation, hybrid, automata, modelling.
\end{IEEEkeywords}

%
\IEEEpeerreviewmaketitle

\section{Introduction}
\label{sec:intro}

The human heart is a vital organ and is responsible for pumping blood
around the body to other vital organs.  Patients can develop abnormal
cardiac behaviour, such as bradycardia (slow heart rate). Cardiac
pacemakers can treat bradycardia by monitoring the patient's heart and
delivering electrical stimuli to the heart when needed. Pacemakers are
life-critical medical devices that must be certified against stringent
safety standards, such as IEC~60601-1~\cite{iec60601}. Certification is
a costly and time consuming process, yet 1,210 computer-related recalls
for medical devices were reported to the US Food and Drug Administration
between 2006 and 2011~\cite{AlemzadehIKR13}.

Pacemakers must be validated by clinical trials as part of the
certification process. This requires the pacemaker to be tested in
closed-loop with a patient's heart. Since clinical trials are the only
times when a pacemaker is tested on a real heart, they provide a glimpse
of how well the pacemaker performs in the real world. Clinical trials
are usually performed late in the product development phase, because they
are costly and time consuming to manage. Thus, issues found during a
clinical trial may be costly and time consuming to fix and a new
clinical trial may be required to re-evaluate the pacemaker. Some
limitations of clinical trials include: potentially small sample of
patients that are not representative of the general population,
difficulty in recruiting patients with specific heart conditions,
difficulty in interrogating a patient's heart to better understand
design issues, and inherent risk to the patients.

Recently, the emulation of the heart has been proposed to facilitate the
closed-loop testing of pacemakers~\cite{JiangPM12}. Emulation is the
real-time simulation of a heart model that can react to a pacemaker's
electrical shocks and also output the heart's electrical activities for
the pacemaker to sense. High-fidelity heart models provide realistic
behaviour but are computationally
intensive~\cite{BordasCFMNPJ09,BartocciCGGSF11}, thus, precluding them
from emulation. The following benefits can be gained if high-fidelity 
heart models can be emulated: cheaper and quicker testing than with
clinical trials, earlier testing of pacemakers in closed-loop 
in the development phase and outside of clinics, greater testing 
coverage by emulating a range of heart conditions, better understanding
of design issues by interrogating the emulated heart (e.g.,
replaying problematic test cases), and having minimal risk to the patients. We
envision the use of emulated hearts alongside clinical trials to help
accelerate the certification process.

In the context of testing cardiac pacemakers, a heart model should
possess the following properties:
\begin{itemize}
\item \textbf{Abstraction:} The model focusses on the important aspects by
  ignoring irrelevant details. For example, the cardiac conduction
  system is the most important aspect because it is responsible for
  coordinating the heart's electrical activities. Irrelevant details
  may include hemodynamics (e.g., blood flow), mechanics (e.g., muscle
  movement), and chemistry (e.g., cellular reactions).

\item \textbf{Accuracy:} The model faithfully represents the cardiac conduction
  system and demonstrates realistic behaviours. A high-fidelity model
  provides an accurate reflection of reality but requires high
  computational power. A lower fidelity model requires less
  computational power but at the risk of providing an inaccurate
  reflection of reality.
  
\item \textbf{Prediction:} The model can answer questions about a real heart,
  such as ``How does the heart respond when setting X of the pacemaker
  is used?''
  
\item \textbf{Inexpensiveness:} The model should be cheaper and faster to construct
  and use the emulated heart than to conduct a clinical trial.
\end{itemize}
The heart models of Chen et al.~\cite{ChenDKM14}, Jiang et
al.~\cite{JiangPAM14}, and M{\'e}ry and Singh~\cite{MeryS12} consider
just the emergent features of the cardiac conduction system, which is
composed of millions of cells.  They model the conduction system as a
static, two-dimensional, sparse network of cardiac cells. Jiang et
al.~\cite{JiangPM12} also developed a hardware prototype that emulates
the cardiac conduction system as discrete events. When the logic of a
pacemaker's software is tested in closed-loop with a heart model, it may
be sufficient to use a heart model that produces and responds to
discrete events~\cite{JiangPAM14,MeryS12}. However, when a physical
implementation of the pacemaker is tested in closed-loop with a heart
model, it is necessary to use a heart model that produces and responds
to continuous-time signals. This is because the physical pacemaker
expects a real heart as its environment. The heart model of Chen et
al.~\cite{ChenDKM14} simulates the conduction system as continuous-time
signals. However, the signals are too abstract and bear little
resemblance with reality. Thus, the model lacks the accuracy and,
therefore, the predictive power.

\subsection{Contributions}
This paper reviews the state-of-the-art heart
models~\cite{ChenDKM14,JiangPM12,JiangPAM14,MeryS12,LianKM10} that have
been designed specifically for the closed-loop testing of cardiac
pacemakers. Without introducing significant computational complexity, we
propose significant improvements to the modelling of the cardiac
conduction system to create a heart model that produces realistic
continuous-time electrical signals that a pacemaker would sense. Our
model faithfully models forward and backward conduction, which is
essential in the modelling of complex re-entrant
circuits~\cite{KleberR04,Spector13,ManiP14} that cause arrhythmia
(abnormal heart rate). Our primary contributions are:
\begin{itemize}
\item We develop a continuous-time model of the conduction system as a
  two-dimensional network of cardiac cells. Each cell produces an
  accurate continuous-time signal that represents its electrical
  activities. These signals are propagated continuously along the
  paths between the cells. Complex conduction behaviours, such as
  arrhythmias caused by re-entrant circuits, can be reproduced
  faithfully by our model. Our heart model is easily customised by
  modifying various parameters of the conduction system.
  
\item Each cardiac cell in our heart model is based on the hybrid
  automaton developed by Ye et al.~\cite{YeESG08}. We have greatly
  improved the design of the hybrid automaton to overcome the following
  limitations: the cell becomes unstable when it is stimulated in quick
  succession, and the cell is too sensitive to electrical stimulation
  from its neighbours. Our improvements are elaborated in
  Section~\ref{sec:cell}.
  
\item Each path in our heart model is modelled with timed automata.  The
  path model is inspired by that of Jiang et al.~\cite{JiangPAM14} that
  was designed to propagate discrete events rather than continuous-time
  signals. Our path model is elaborated in Section~\ref{sec:path}.
  
\item We demonstrate in Section~\ref{sec:results} that a
  MathWorks\textsuperscript{\textregistered}
  Simulink\textsuperscript{\textregistered} and
  Stateflow\textsuperscript{\textregistered} implementation of our heart
  model can simulate a wide range of heart conditions with realistic
  results.
\end{itemize}

\subsection{Paper Layout}
Section~\ref{sec:background} provides a background to the cardiac
conduction system and the important features of electrical activities
that are sensed by a pacemaker.  Section~\ref{sec:related} reviews the
state-of-the-art heart models for closed-loop testing of pacemakers.
Section~\ref{sec:cell} reviews the computationally efficient hybrid
automata model of cardiac cells developed by Ye et
al.~\cite{YeESG08}. We identify the limitations encountered with Ye et
al.'s model during simulation and how we corrected them.
Section~\ref{sec:path} describes our path model that handles the
propagation of continuous-time signals.  In Section~\ref{sec:heart}, we
create our proposed heart model by composing instances of our cardiac
cell and path models into a network that replicates the conduction
pathway.  Section~\ref{sec:results} evaluates the capabilities of our
proposed heart model with the recent heart model of Chen et
al.~\cite{ChenDKM14}.  Section~\ref{sec:conclusion} concludes this paper
and discusses future work for improving the proposed heart model.


\section{Background}
\label{sec:background}

\begin{figure}
	\centering

	\includegraphics[width=0.9\columnwidth]{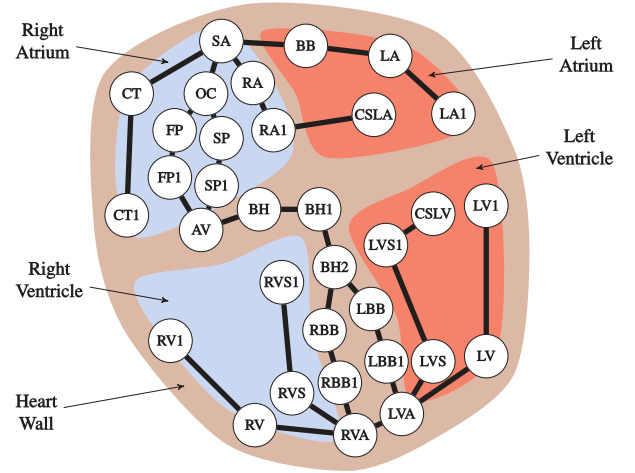}
	
	\caption{Schematic of the heart and conduction system.}
	\label{fig:background:heart}
\end{figure}

The heart pumps blood around the body in a rhythmic manner.
Figure~\ref{fig:background:heart} is a schematic of the heart
and shows its four chambers: the right and left
atriums and ventricles. The right atrium and ventricle are
responsible for pumping deoxygenated blood through the
lungs, while the left atrium and ventricle are responsible
for pumping oxygenated blood through the body. The
contractions of the chambers are coordinated by electrical
stimuli that propagate throughout the heart's conduction
system. The conduction pathways are shown in
Figure~\ref{fig:background:heart} as solid black lines with
dots at important locations. The names of these locations
are labelled with an acronym and their full forms are given
in Table~\ref{tab:background:heart}. 

\begin{table}
	\centering
	\renewcommand{\arraystretch}{1.3}
	\caption{Full names of nodes along the conduction pathways.}
	\label{tab:background:heart}
	
	\begin{tabular}{l l c l l}
		\hline
		AV	& Atrioventricular		&	& LVS	& Left ventricular septum	\\
		BB	& Bachmann's bundle		&	& OC	& Os cordis					\\
		BH	& Bundle of His			&	& RA	& Right atrium				\\
		CS	& Coronary sinus		&	& RBB	& Right bundle branch		\\
		CT	& Crista terminalis		&	& RV	& Right ventricle			\\
		FP	& Fast path				&	& RVA	& Right ventricular apex	\\
		LA	& Left atrium			&	& RVS	& Right ventricular septum	\\
		LBB	& Left bundle branch	&	& SA	& Sinoatrial				\\
		LV	& Left ventricle		&	& SP	& Slow path					\\
		LVA	& Left ventricular apex	&	&		&							\\
		\hline
	\end{tabular}
\end{table}

\subsection{Cardiac Cycle}

\begin{figure}
  \centering
  
  \subfloat[Phase 1]{
    \includegraphics[width=0.4\columnwidth]{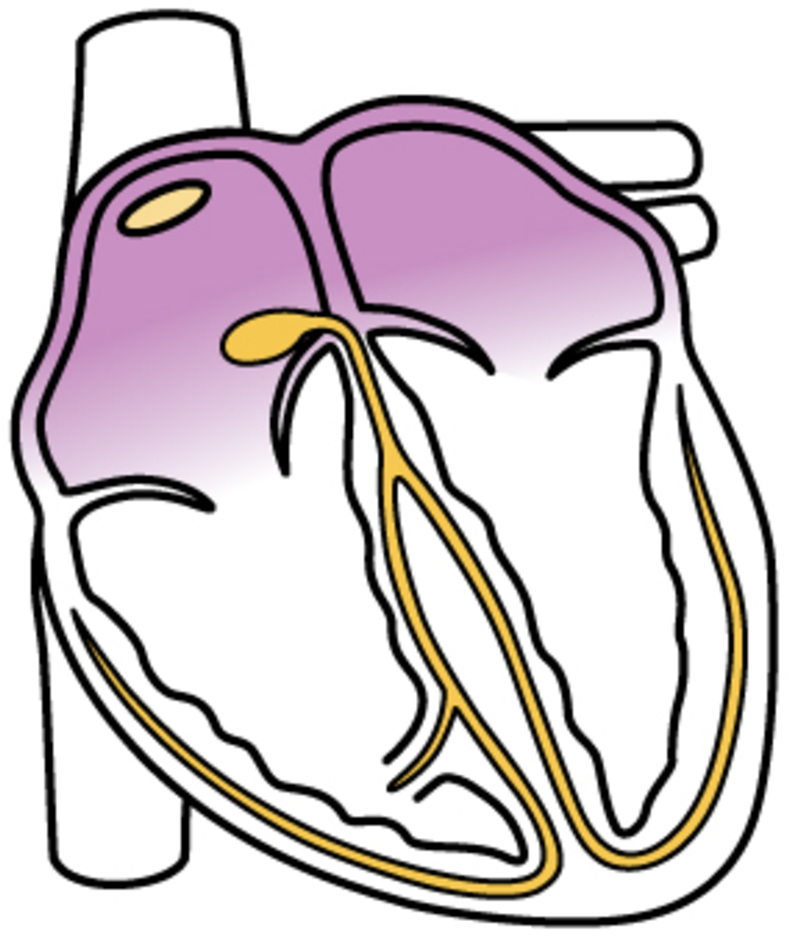}
    \label{fig:background:cycle_1}
  }
  \hfill
  \subfloat[Phase 2]{
    \includegraphics[width=0.4\columnwidth]{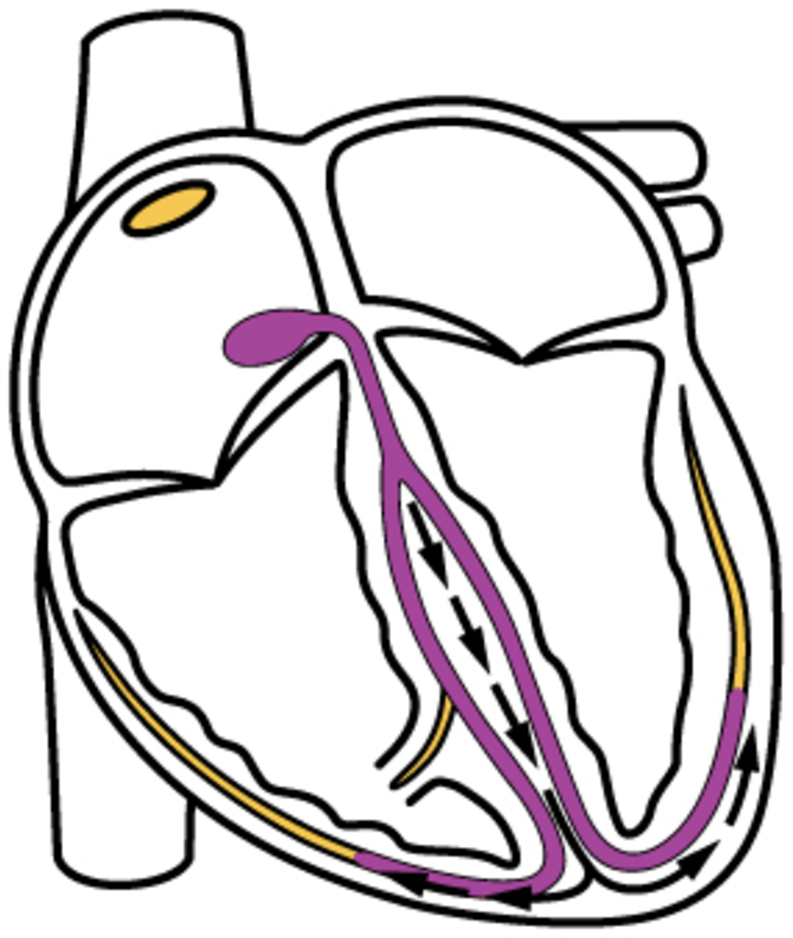}
    \label{fig:background:cycle_2}
  }
  
  \subfloat[Phase 3]{
    \includegraphics[width=0.4\columnwidth]{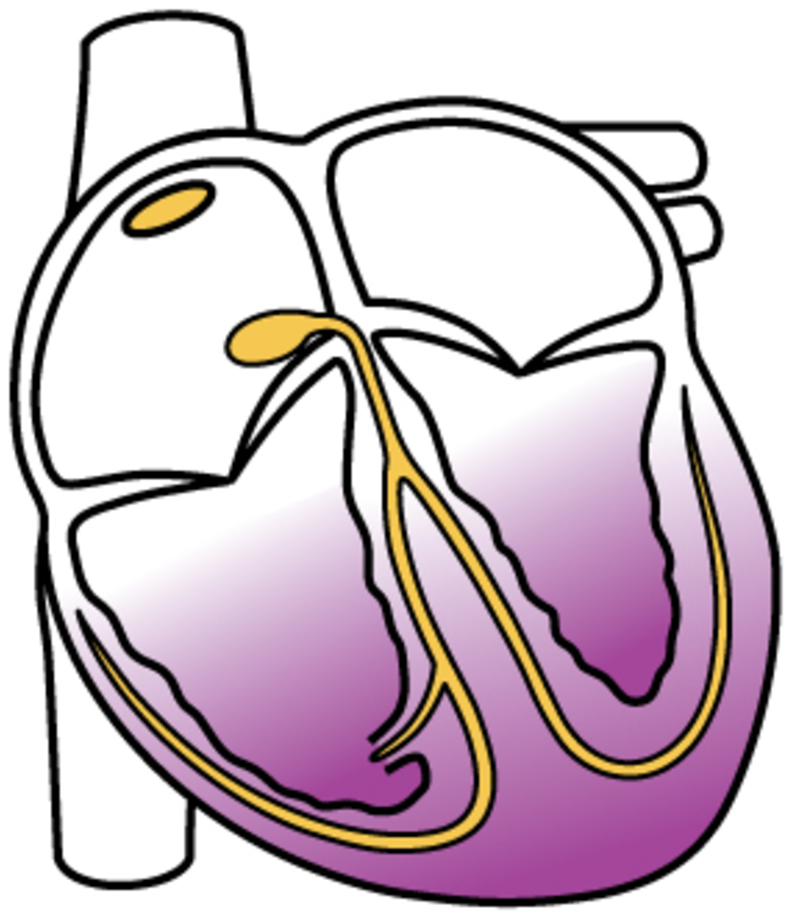}
    \label{fig:background:cycle_3}
  }
  \hfill
  \subfloat[Phase 4]{
    \includegraphics[width=0.4\columnwidth]{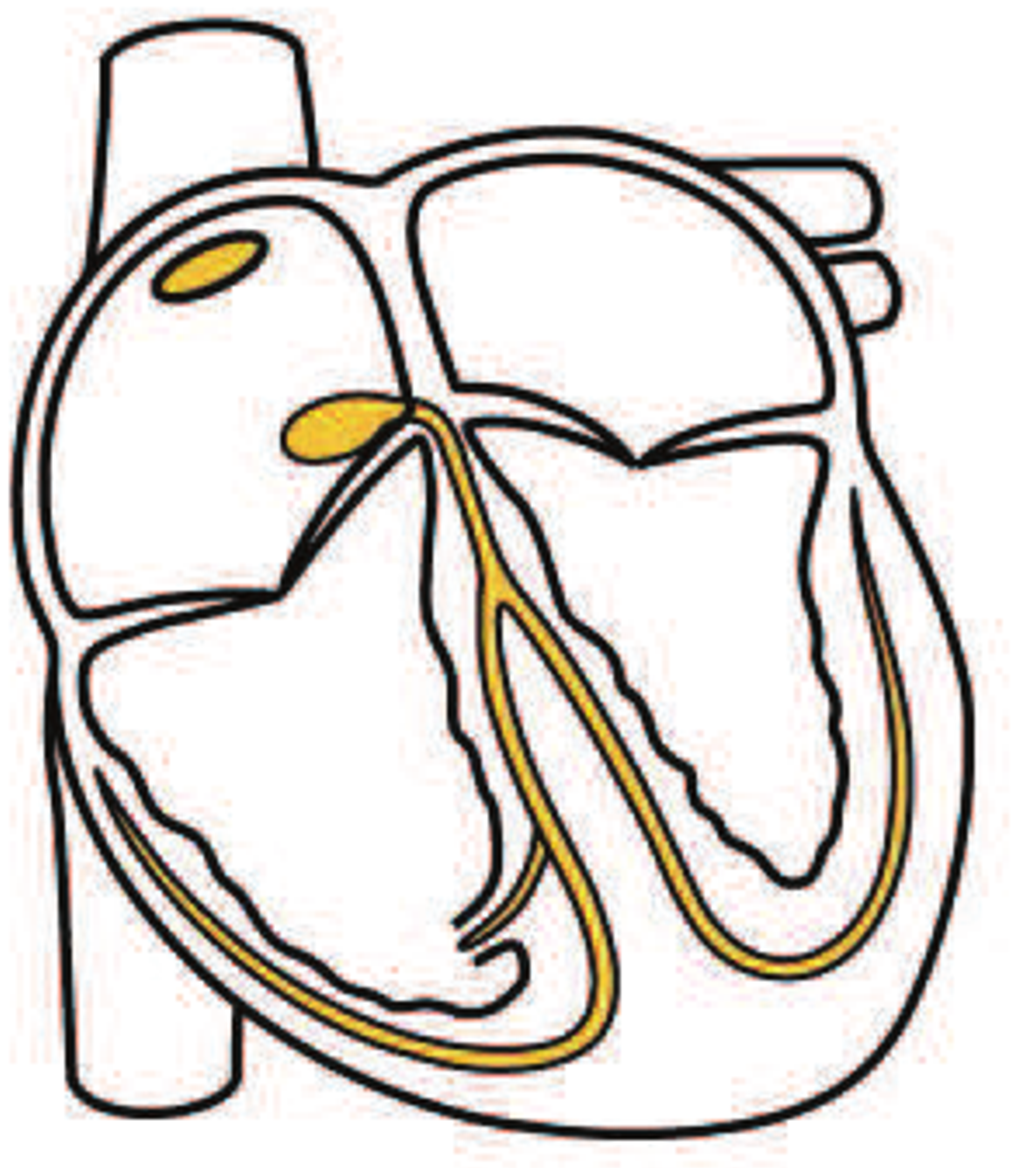}
    \label{fig:background:cycle_4}
  }
  
  \caption{Phases of the cardiac cycle. Adapted from http://philschatz.com/ anatomy-book/contents/m46664.html\#sinoatrial-sa-node.}
  \label{fig:background:cycle}
\end{figure}

This section describes the major actions of the heart during
one cardiac cycle (one heart beat) with the help of
Figure~\ref{fig:background:cycle}. In the first phase of the
cardiac cycle, Figure~\ref{fig:background:cycle_1}, the
sinoatrial (SA) node generates an electrical stimulus that
spreads quickly throughout the right and left atriums. This
causes the atriums to contract, pumping blood from the
atriums into the ventricles. In the second phase,
Figure~\ref{fig:background:cycle_2}, the electrical stimulus
reaches the atrioventricular (AV) node and is delayed
momentarily before it continues down into the ventricles.
This delay is very important because it gives the atriums
enough time to contract and fully fill the (relaxed)
ventricles. In the third phase,
Figure~\ref{fig:background:cycle_3}, the electrical stimulus
reaches the right and left ventricular apexes and travels
out to the fast conducting Purkinje fibers. This causes the
right and left ventricles to contract and pump out blood. In
the fourth phase, Figure~\ref{fig:background:cycle_4}, the
ventricles relax after pumping out all their blood.

In a normal heart, each cardiac cycle begins from the SA
node, which generates periodic electrical stimuli that
spread through the conduction system. The following sections
describe the genesis of the heart's electrical activity and
how it appears to a pacemaker. Finally, we describe some
common arrhythmias that a heart model should aim to capture.

\subsection{Action Potentials of Cardiac Cells}
\label{sec:background:ap}

\begin{figure}
  \centering
  
  \includegraphics[width=\columnwidth]{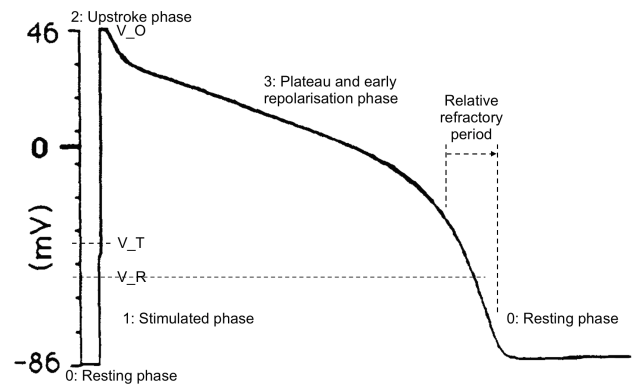}
  
  \caption{Phases of the action potential. Adapted from~\cite{LuoR94}.}
  \label{fig:background:ap}
\end{figure}

Most of the heart's electrical activities, that a pacemaker 
senses, are generated by the myocytes (muscle cells)~\cite{BunchHSAF13}. 
A cell's electrical activities result from the
movement of ions across its membrane, creating potential
differences. The cell's electrical response to an electrical
stimulus is described by its \emph{action potential}~\cite{LuoR91}. 
Figure~\ref{fig:background:ap}
shows the four phases of a typical action potential, which
plots the cell's membrane potential over time. In the
\textbf{resting phase}, the cell is inactive and has a resting
potential of approximately $-85mV$. The cell enters
the \textbf{stimulated phase} when excited electrically
by its neighbours or by an artificial pacemaker. The cell
returns to the resting phase if its membrane potential fails
to cross the threshold voltage $V_T$ of approximately
$-40mV$ when the excitation stops. Otherwise, the cell
enters the \textbf{upstroke phase} and \emph{depolarises} by
allowing ions to move rapidly across its membrane, causing
its membrane potential to reach an overshoot voltage $V_O$ of
approximately $+45mV$. Then the cell enters the
\textbf{plateau and early repolarisation phase}. The cell
contracts and starts to \emph{repolarise}, i.e., its
membrane potential starts to return to its resting potential. 
When the membrane potential is less than the voltage $V_R$ of
approximately $-55mV$, the cell has relaxed and returned
to the resting phase.

\begin{figure}
  \centering
  
  \includegraphics[width=\columnwidth]{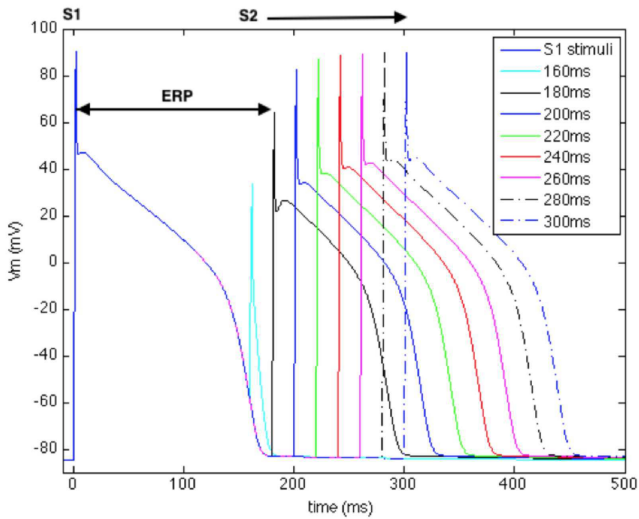}
  
  \caption{Dynamic behaviour of secondary excitations. Adapted from~\cite{Ou15}.}
  \label{fig:background:secondary}
\end{figure}

All cardiac cells can only respond to subsequent excitations in the
later portion of its action potential, called the
\emph{relative refractory period}. However, the membrane
potential must cross a higher threshold voltage.
Figure~\ref{fig:background:secondary} shows a normal action
potential at $0ms$ and some possible secondary excitations
between $160$ and $300ms$. For a secondary excitation at
$180ms$, the resulting action potential has a lower
overshoot voltage $V_O$ and a shorter action potential
duration. The secondary excitation at $300ms$ results in a
more normal action potential because the cell has
rested for a longer period.


Prominent biophysical cardiac cell models, which explain the
genesis of action potentials in terms of ionic flow, include
Luo-Rudy~\cite{LuoR91} and Hodgkin-Huxley~\cite{HodgkinH52}.
Although biophysical models have high-fidelity, they are
computationally intensive. Ye et al.~\cite{YeEGS05,YeESG08}
create computationally efficient cardiac cell models by
considering just the emergent features of the biophysical
models, i.e., the action potential and its dynamic
response to secondary excitation. It should be noted that
the action potential duration of a human ventricular myocyte 
is approximately twice that of an atrial myocyte.


\subsection{Action Potentials and the Electrogram}
\label{sec:background:egm}

%
%

\begin{figure}
	\centering

	\subfloat[Three action potentials (APs) from different regions of the heart and their corresponding electrograms (EGMs). Adapted from~\cite{HawsL90}.]{
		\includegraphics[width=0.9\columnwidth]{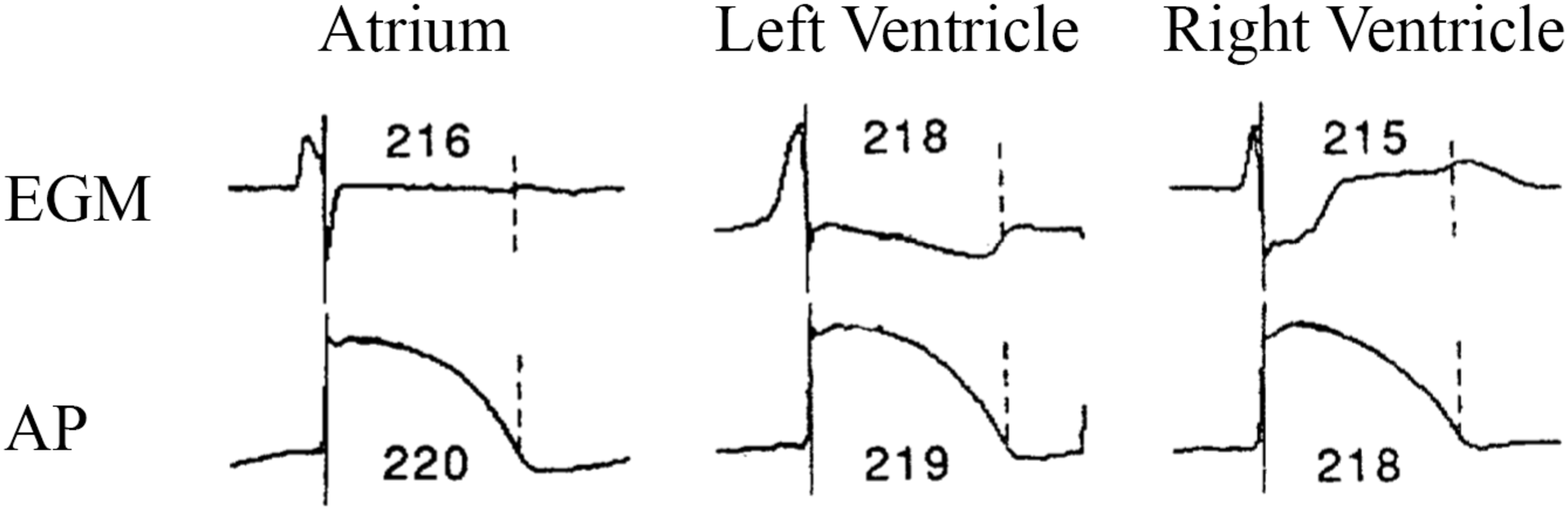}
		\label{fig:background:egm_ap}
	}

	\subfloat[EGM deflections due to a travelling electrical wavefront. Adapted from~\cite{BunchHSAF13}.]{
		\includegraphics[width=0.7\columnwidth]{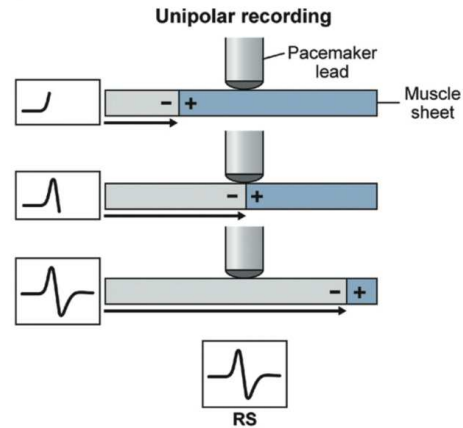}
		\label{fig:background:lead}
	}

	\caption{Electrograms (EGMs).}
	\label{fig:background:egm}
\end{figure}

A key function of any pacemaker is to sense the heart's
electrical activity, by using one or more electrodes 
attached to the inside of the heart wall. The electrical 
activity of the cardiac cells in the electrode's 
immediate vicinity are sensed most strongly. A recording
of the sensed activities is called an electrogram 
(EGM)~\cite{Kusumoto10}. Figure~\ref{fig:background:egm_ap} 
shows three action potentials and their corresponding 
EGMs. To help understand the EGM, Figure~\ref{fig:background:lead}
shows that the EGM deflects up and down whenever an
electrical wavefront passes under the electrodes~\cite{BunchHSAF13}. 
The faster that the wavefront passes, the steeper the deflection.
In Figure~\ref{fig:background:egm_ap}, two distinct deflections 
can be seen in each EGM and they correspond with the upstroke
and resting phases of their respective action potential.


A heart
model that produces distorted action potentials will also
produce distorted EGMs. Such distorted EGMs cannot be used
to reliably test a pacemaker's ability to discern the timing
of important cardiac activities. Moreover, the predictive
power of a heart model is compromised when the action
potentials are distorted. For example, a heart model with
accurate action potentials might predict that a cell goes
into its upstroke phase because its neighbours' voltages are
high enough. However, a heart model with distorted action
potentials might instead predict that the cell returns to
its resting phase because its neighbours' voltages are too
low. Thus, arrhythmia would be predicted incorrectly. 

\begin{table*}
  \centering
  \renewcommand{\arraystretch}{1.3}
  \caption{Qualitative comparison of heart models that are designed for testing cardiac medical devices. AP = action potential. HA = hybrid automata. TA = timed automata.}
  \label{tab:related}
  
  \begin{tabular}{| L{1cm} | C{2cm} | C{2cm} | C{2cm} | C{2cm} | C{2cm} | C{2cm} | C{2cm} |}
    \cline{2-8}
    \multicolumn{1}{L{1cm}|}{}   	& {\bf Reality}                                             			& \multicolumn{6}{C{14.2cm}|}{Less Abstract $\leftarrow$ \hfill {\bf Heart Models} \hfill $\rightarrow$ More Abstract}                                                                                                                                                                                                                                                                                    						\\
    \cline{2-8}
    \multicolumn{1}{L{1cm}|}{}		& {\bf Real Heart}~\cite{DurrerDFJMA70}                                 & {\bf Hi-Fi}~\cite{SermesantDA06,SmaillH10}                  							& {\bf UoA}                                                     & {\bf Oxford}~\cite{ChenDKM14}                       			& {\bf UPenn}~\cite{JiangPM12,JiangPAM14}	& {\bf LORIA}~\cite{MeryS12}                        & {\bf MES}~\cite{LianKM10}                                                               				\\ \hline
    {\bf Cell Model}				& Continuous APs from biophysical processes~\cite{ClaytonBCDFMPSSZ11}	& Continuous APs from biophysical models~\cite{ClaytonBCDFMPSSZ11}						& Continuous APs from improved Stony Brook HA~\cite{YeESG08}	& Continuous APs from simplified Stony Brook HA~\cite{YeEGS05}	& Discrete APs from TA   					& Discrete APs from logico-mathematics				& \multirow{2}{2cm}{\centering Continuous AV signal generators mimic whole heart electrical activity}	\\ \cline{1-7}
    {\bf Path Model}				& Continuous propagations from biophysical processes~\cite{KleberR04}	& Continuous propagations from reaction-diffusion equations~\cite{ClaytonBCDFMPSSZ11}   & Continuous propagations from TA and contribution function		& Continuous propagations from contribution function  			& Discrete propagations from TA				& Discrete propagations from cellular automata		& 																										\\ \hline
    {\bf Spatial Model}    			& 3D tissue (layers of bundles of fibers) that deforms       			& 3D finite-volume that deforms              	              							& \multicolumn{4}{C{9.2cm}|}{2D, static, and sparse network of cells along the conduction pathway}                                                                                                                           	& Black boxes of major heart components                													\\ \hline
  \end{tabular}
\end{table*}

\subsection{Common Arrhythmias}
\label{sec:background:arrhythmia}

Arrhythmias can be caused by abnormalities in the generation and
propagation of action potentials through the conduction system. The
abnormalities may be due to congenital defects, side-effects of
medication, or cell death. The following are some common
arrhythmias~\cite{Kusumoto10,Spector13,ManiP14} that a heart model for
pacemaker testing should aim to capture:
\begin{itemize}
\item \textbf{Heart block:} This occurs when electrical stimuli has
  difficulty propagating through the AV node. The propagation of the
  stimuli may be delayed for longer than usual or may be prevented from
  propagating altogether.
  
\item \textbf{AV node re-entrant tachycardia:} This occurs when a
  re-entry circuit forms around the AV node, causing tachycardia.
  
\item \textbf{Bundle branch block:} This occurs when electrical stimuli
  travels slower or not at all down one of the bundle branches.
  
\item \textbf{Wolff-Parkinson-White syndrome:} This occurs when there is
  an extra conduction pathway between the atriums and ventricles. The
  extra pathway allows electrical stimuli to bypass the AV node and
  create a feedback loop between the atriums and ventricles.
  
\item \textbf{Long Q-T syndrome:} This occurs when the repolarisation of
  the ventricles is delayed, i.e., their action potential durations are
  longer than usual.
  
\item \textbf{VA conduction:} This occurs when electrical stimuli from
  the ventricles conduct backwards through the conduction pathways and
  into the atriums.
\end{itemize}

Pacemakers can also cause arrhythmias when they are unable to correctly
sense the timing of the heart's electrical activities. For example,
pacemaker-mediated tachycardia is caused by the pacemaker inadvertently
conducting electrical stimuli from the ventricles back to the
atriums. Pacemakers that deliver electrical stimuli that are not
synchronised with the heart's cardiac rhythm can cause the heart to
fibrillate~\cite{McLeodJ04}, i.e., twitch uncontrollably.


\section{Related Work}
\label{sec:related}

The electrophysiology of the heart has been well
researched~\cite{KleberR04,FranzonePS14}, resulting in the proposal of
many theories. These theories are validated by creating high-fidelity
whole heart models~\cite{SermesantDA06,SmaillH10} and ascertaining if
they can reproduce experimental observations, i.e., the models are
accurate, realistic, and predictive. These high-fidelity models are
useful in predicting the prognosis of patient-specific heart
conditions~\cite{Trayanova14} and in assisting with interventional
cardiology~\cite{GreenKPRSLMM14}.

On the other hand, abstract heart models have been developed with the
goal of enabling the closed-loop testing of cardiac
pacemakers. Table~\ref{tab:related} provides a qualitative comparison of
existing heart models. The abstract heart models from
Oxford~\cite{ChenDKM14}, UPenn~\cite{JiangPM12,JiangPAM14},
LORIA~\cite{MeryS12}, and MES~\cite{LianKM10} are designed for testing
the pacemaker logic. To enable the formal verification of the pacemaker
logic, Oxford, UPenn, and LORIA use hybrid automata (HA) or timed
automata (TA) to develop formal models of the cardiac conduction
system. UPenn and LORIA model the transitions between the resting,
upstroke, and early refractory phases of the action potential as
discrete events on a continuous timeline. These discrete events are
propagated between cells and the propagation is either successful or
unsuccessful. These abstractions result in heart models that may produce
more behaviours than is possible by a real heart, i.e., an
over-approximation. Thus, all problems detected during closed-loop
testing must be validated against a more concrete heart
model~\cite{JiangPAM14}.

The heart model from Oxford~\cite{ChenDKM14} is more concrete than those
from UPenn and LORIA because Oxford models the action potentials as
continuous signals. Oxford uses a simplified model of the cardiac cell
that Ye et al.~\cite{YeEGS05} developed with hybrid automata. Oxford
incorporates a $g(\vec{v})$ function into the cell model to capture the
continuous electrical activity that a cell receives from its
neighbours. However, the $g(\vec{v})$ function does not consider the
directional behaviour of electrical propagation due to the refractory
period of cardiac cells. Noting these limitations,
Sections~\ref{sec:cell} and \ref{sec:path} describe our improvements to
the cell and path models. Our heart model can faithfully simulate
complex re-entrant circuits without a significant increase in
computational complexity.


\section{Cardiac Cell Model}
\label{sec:cell}

\begin{figure}		
	\includegraphics[width=0.9\columnwidth]{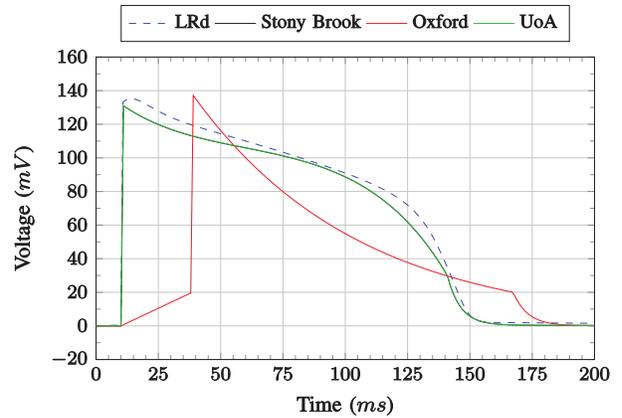}
	
	\caption{Comparison of the action potentials (APs) produced by Luo-Rudy~\cite{LuoR91}, 
			 Stony Brook~\cite{YeESG08}, Oxford~\cite{ChenDKM14}, and our improved version (UoA). 
			 Note that the AP of UoA overlaps that of Stony Brook's because they are identical.}
	\label{fig:cell:ap_comparison}
\end{figure}

The Oxford heart model~\cite{ChenDKM14} uses a simplified version of the
isolated cardiac cell model developed by Stony Brook~\cite{YeEGS05}. The
Stony Brook model is itself a simplification of the Luo-Rudy
model~\cite{LuoR91} because it only models the action potential and its
dynamic response to secondary excitation. The Stony Brook model uses
three piecewise-continuous variables, called $v_x$, $v_y$, and $v_z$, to
capture different features of the action potential. The sum of these
three variables produces the action potential. The Oxford heart model
discards the $v_y$ and $v_z$ variables and retains just the $v_x$
variable.  Unfortunately, Figure~\ref{fig:cell:ap_comparison} shows that
the resulting action potential is no longer realistic and this
diminishes the predictive power of Oxford's heart model
(Section~\ref{sec:background:egm}). Thus, for our heart model, we retain
all three variables and we use an updated Stony Brook cardiac cell
model~\cite{YeESG08}. This section describes the Stony Brook model in
more detail and our improvements that overcome some of the model's
limitations.

\subsection{Stony Brook Cardiac Cell Model}

\begin{figure}
  \centering
 \includegraphics[width=0.9\columnwidth]{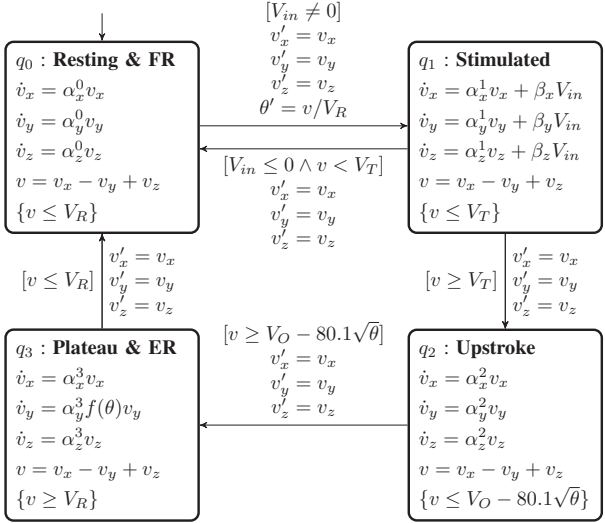}
  
  \caption{Stony Brook cardiac cell model~\cite{YeESG08}.}
  \label{fig:cell:stony_brook}
\end{figure}

\begin{table}
  \centering
  \renewcommand{\arraystretch}{1.5}
  \caption{Coefficients and constants in the Stony Brook cardiac cell model~\cite{YeESG08}.}
  \label{tab:cell:stony_brook}
  
  \begin{tabular}{l l l l}
    \hline
    $\alpha^0_x = -0.0087$	& $\alpha^1_x = -0.0236$	& $\alpha^2_x = -0.0069$	& $\alpha^3_x = -0.0332$	\\
    $\alpha^0_y = -0.1909$	& $\alpha^1_y = -0.0455$	& $\alpha^2_y =  0.0759$	& $\alpha^3_y = 0.0280$		\\
    $\alpha^0_z = -0.1904$	& $\alpha^1_z = -0.0129$	& $\alpha^2_z =  6.8265$	& $\alpha^3_z = 0.0020$		\\
    \hline
    $V_R = 30$				& $\beta_x = 0.7772$		&							&							\\
    $V_T = 44.5$			& $\beta_y = 0.0589$		&							&							\\
    $V_O = 131.1$			& $\beta_z = 0.2766$		&							&							\\
    \hline
  \end{tabular}
\end{table}

The Stony Brook model~\cite{YeESG08} models the time course
of the action potential with the \emph{hybrid automaton} (HA) shown
in Figure~\ref{fig:cell:stony_brook}. The four phases of the
action potential, described in
Section~\ref{sec:background:ap}, are represented as four
locations in the HA. In each location, the membrane
potential is defined by the variable $v$ as the sum of the
variables $v_x$, $v_y$, and $v_z$. The rates at which the
variables $v_x$, $v_y$, and $v_z$ change are defined by
their derivatives $\dot{v}_x$, $\dot{v}_y$, and $\dot{v}_z$,
respectively. The values of the coefficients and constants
are given in Table~\ref{tab:cell:stony_brook}. Note that the
Stony Brook model offsets all the voltages such that the
resting potential is at $0mV$.

By default, the HA always starts in location $q_0$, which is
the resting phase of the action potential. It stays in $q_0$
as long as the invariant $v \le V_R$ is true. The HA
transitions from location~$q_0$ to~$q_1$ when the
voltage~$V_{in}$ around the cell is greater than $0mV$.
During this transition, the last values of $v_x$, $v_y$, and
$v_z$ when leaving~$q_0$ are used to set their initial
values when entering~$q_1$. Recall from
Figure~\ref{fig:background:secondary} that the amplitude and
duration of the action potential depends on how long the
cell had been in the resting phase when it is stimulated. 
This is approximated by assuming that the closer the cell's 
membrane potential is to $0mV$, the longer the cell has been
in location~$q_0$. Thus, the time that the cell
had been in the resting phase is approximated by 
normalising the membrane potential~$v$ against the 
voltage~$V_R$, i.e., $\theta = v/V_R$. In location~$q_1$,
the rate at which the cell's membrane potential increases
depends on the strength of~$V_{in}$. The HA transitions back
to location~$q_0$ if the voltage~$V_{in}$ around the cell
fails to stimulate the cell above the threshold
voltage~$V_T$. However, if the cell's membrane voltage~$v$
is stimulated above~$V_T$, then the HA transitions to 
location~$q_2$. The amplitude of the cell's membrane
potential is calculated as $V_O - 80.1 \sqrt{\theta}$, i.e.,
it depends on how long the cell had been in the resting
phase. The HA transitions to location~$q_3$ and the cell's 
membrane potential
starts to drop. The rate of repolarisation is determined by
the function $f(\theta)$ and depends on how long the cell
had been in the resting phase:
\begin{equation}
	\label{eq:cell:f_theta}
	f(\theta) = 0.29e^{62.89\theta} + 0.70e^{-10.99\theta}
\end{equation}
A higher value for $f(\theta)$ means a faster rate of
repolarisation. Once the cell's membrane voltage~$v$ 
drops below the resting voltage~$V_R$, the HA transitions
back to~$q_0$.

\subsection{Improvement: Stabilising the Action Potential} 
\label{sec:cell:improvement_1}

\begin{figure}
	\centering
	\includegraphics[width=0.9\columnwidth]{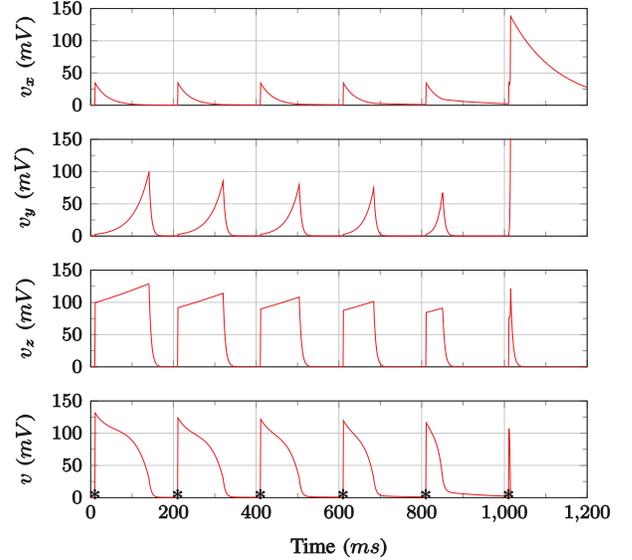}

	\caption{Action potentials shorten unnaturally for the Stony Brook model. Stimulated at $200ms$ intervals}
	\label{fig:cell:improvement_1a}
\end{figure}

\begin{figure}
	\centering
	\includegraphics[width=0.9\columnwidth]{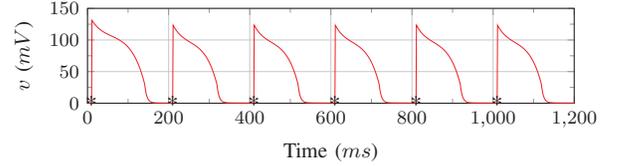}

	\caption{Action potentials for the UoA model when stimulated at $200ms$ intervals.}
	\label{fig:cell:improvement_1b}
\end{figure}

The bottom plot in Figure~\ref{fig:cell:improvement_1a}
shows the action potentials produced by the Stony Brook model
when stimulated at $200ms$ intervals. We can see that the
duration of the action potentials shorten towards zero over time.
This is unnatural because the action potentials
should settle to a constant duration when stimulated at a
constant interval~\cite{NolascoD68}. The top three plots in
Figure~\ref{fig:cell:improvement_1a} show the values of
$v_x$, $v_y$, and $v_z$, respectively, over time. 
Note that $v_x$ describes the initial voltage drop of the
plateau phase. We can see
that the~$v_x$ variable takes too long to decrease in location~$q_0$. 
Thus, each time the cell is stimulated, it
enters location~$q_1$ with a slightly higher
value for~$v_x$. This causes the values of~$\theta$ and
$f(\theta)$ to increase over time. An increasing value of
$f(\theta)$ causes a faster rate of repolarisation and,
hence, shorter action potential durations.

\begin{figure}

	\centering
	\includegraphics[width=0.9\columnwidth]{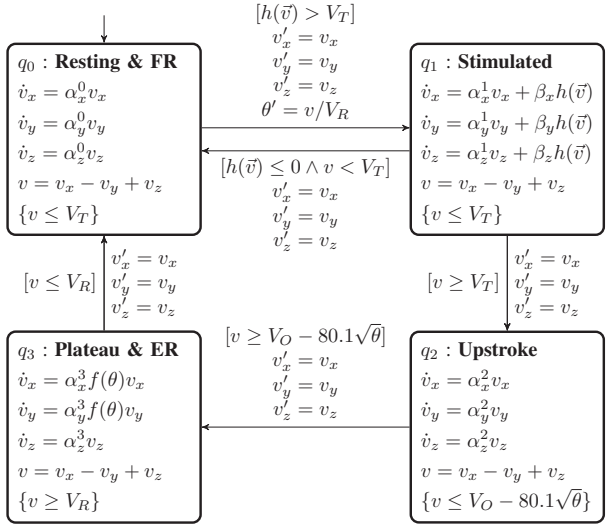}
  
  \caption{Improved cardiac cell model (UoA).}
  \label{fig:cell:improved}
\end{figure}

To prevent the unnatural shortening of the action 
potential, the value of~$v_x$ needs to be closer to 
zero when the cell transitions from location~$q_3$
to~$q_0$. Thus, we increase the rate at which~$v_x$ 
decreases towards zero by including the function 
$f(\theta)$ in location~$q_3$ for~$\dot{v}_x$. 
The improved HA is shown in Figure~\ref{fig:cell:improved}.
Figure~\ref{fig:cell:improvement_1b} shows that the action
potentials produced by the improved HA have constant
durations when stimulated at $200ms$
intervals.

\subsection{Improvement: Bounding the Rate of Repolarisation}
\label{sec:cell:improvement_2}

In the Stony Brook model, the action potential duration
approaches zero if the cell was stimulated shortly after
entering location~$q_0$. This behaviour is unnatural because
the minimum action potential duration that a cardiac cell
can achieve is approximately $40ms$. This problem is due to
the function $f(\theta)$. When the HA returns to
location~$q_0$ from $q_3$, the value of~$v$ is~$V_R$. An
immediate stimulation to location~$q_1$ would set $\theta =
V_R/V_R = 1$. Consequently, $f(1) = 5.96 \times 10^{26}$,
which causes an extremely fast rate of repolarisation.

\begin{figure}		

\centering
\includegraphics[width=0.9\columnwidth]{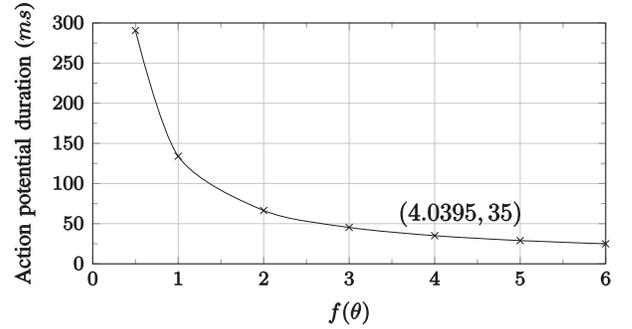}	
	
	\caption{Relationship between the action potential duration (APD) and the value of $f(\theta)$.}
	\label{fig:cell:bounding_f(theta)}	
\end{figure}

To prevent such a fast rate of repolarisation from
occurring, we limit the maximum value that function
$f(\theta)$ can return. Figure~\ref{fig:cell:bounding_f(theta)} 
shows that action potential 
durations of $35~ms$ are produced when $f(\theta = 0.04) = 4.0395$.
Thus, we impose a maximum value of $4.0395$ for the 
function $f(\theta)$:
\begin{equation*}
  f(\theta) = 
  \left \lbrace
    \begin{array}{ll}
      0.29e^{62.89\theta} + 0.70e^{-10.99\theta}	& \text{if}~\theta < 0.04	\\
      4.0395 										& \text{if}~\theta \ge 0.04	\\
    \end{array}
  \right .
\end{equation*}

\subsection{Discussion}
We improved the original Stony Brook cardiac cell
model~\cite{YeESG08} to stabilise the action potentials and
to impose a reasonable minimum action potential duration.
Figure~\ref{fig:cell:improvement_1b} shows that the shape of
the action potentials generated by the improved model
remains realistic. However, it is important to demonstrate
that the action potential durations vary realistically in
response to the timing of secondary excitations (see
Figure~\ref{fig:background:secondary} for illustration).
This dynamic behaviour can be evaluated with a restitution
curve~\cite{NolascoD68}. To plot the restitution curve, the
cardiac cell is electrically stimulated at a constant time
period, called the base cycle length (BCL). The BCL consists
of two time intervals: the action potential duration
followed by the \emph{diastolic interval}. The diastolic
interval begins when the action potential falls to 10\% of
its peak amplitude. For a range of BCLs, the action
potential duration from the tenth BCL is plotted against the
diastolic interval from the ninth BCL.

\begin{figure}
	
	\centering
	\includegraphics[width=0.9\columnwidth]{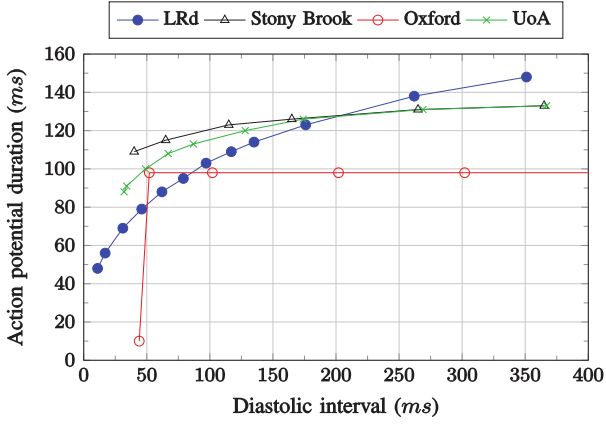}

	\caption{Electrical restitution curves comparing the dynamic behaviour of 
			 Luo-Rudy~\cite{LuoR91}, Stony Brook~\cite{YeESG08}, Oxford~\cite{ChenDKM14}, 
			 and our improved version (UoA).}
	\label{fig:cell:restitution}	
\end{figure}

Figure~\ref{fig:cell:restitution} shows the restitution
curves for the Stony Brook~\cite{YeESG08},
Oxford~\cite{ChenDKM14}, and our improved (UoA) 
models. These models were implemented in MathWorks\textsuperscript{\textregistered}
Simulink\textsuperscript{\textregistered}/Stateflow\textsuperscript{\textregistered}. 
The Oxford implementation was provided by the original
authors~\cite{ChenDKM14}. The restitution curve of the Stony
Brook model has been demonstrated~\cite{YeESG08} to behave realistically
compared to the Luo-Rudy model. However,
because the Stony Brook model is unstable, we could only
reproduce its restitution curve by taking the action
potential duration and diastolic interval of the first
BCL. The restitution curve for our improved model shows
dynamic behaviour that is very similar to Stony Brook. The
difference is due to the changes described in
Section~\ref{sec:cell:improvement_1}. The parameters in the
cell model can be tuned to produce a variety of restitution
curves and is particularly useful when modelling diseased
cardiac cells. The Oxford model, on the other hand, shows
unrealistic behaviour. The action potential duration is
either $9ms$ or $98ms$, depending on whether the diastolic
interval is greater or less than $51ms$. 

\section{Cardiac Path Model}
\label{sec:path}

The Oxford heart model~\cite{ChenDKM14} models the conduction system as
a sparse network of cells (Figure~\ref{fig:background:heart}). The cells
are connected electrically by a voltage contribution function
$g_k(\vec{v})$, which calculates the voltage induced at cell $k$ by its
neighbours:
\begin{equation}
  \label{eq:path:g_k}
  g_k(\vec{v}) = \sum^n_{i = 1} v_i(t - \delta_{ki})a_{ki} - v_k d_k
\end{equation}
For a given cell~$k$ with~$n$ connected neighbours,
$\vec{v} = (v_1 \cdots v_n)$ is a vector of all the neighbours' membrane
potentials, such that~$v_i$ is the membrane potential of
neighbour~$i$. The membrane potential of cell~$k$ is~$v_k$. The time for
cell~$i$'s action potential to propagate to cell~$k$, called the
\emph{conduction time}, is represented by~$\delta_{ki}$.  Thus,
$v_i(t - \delta_{ki})a_{ki}$ represents the membrane potential of cell
$i$ that reaches cell~$k$ after a delay of~$\delta_{ki}$ and with a gain
of~$a_{ki}$.  The strength of cell~$k$'s membrane potential relative to
its neighbours is taken into account by $v_k d_k$, where~$d_k$ is a
distance coefficient.

\begin{figure}
	\centering

	\centering
	\includegraphics[width=0.15\columnwidth]{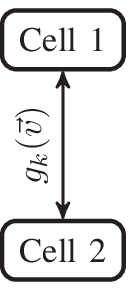}
	\includegraphics[width=0.8\columnwidth]{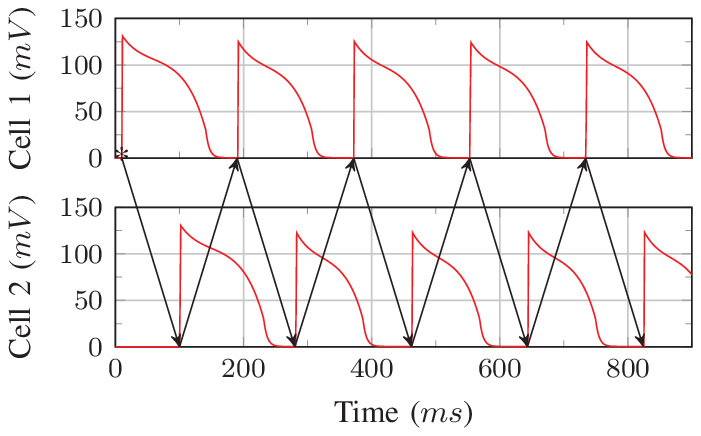}

	\caption{Two cardiac cells connected electrically by the Oxford voltage contribution function $g_k(\vec{v})$.}
	\label{fig:path:oxford}
\end{figure}

Figure~\ref{fig:path:oxford} plots the action potentials of two cardiac
cells connected by Oxford's function $g_k(\vec{v})$. Only cell~$1$
receives an external stimulus at~$10ms$ and the time delay between
the cells is $90ms$. The expected behaviour is for
cell~$1$ to produce an action potential that stimulates cell~$2$ to
produce an action potential. Cell~$2$'s action
potential would not propagate back to cell~$1$ because of the refractory
feature of cardiac cells~\cite{Spector13}. However,
Figure~\ref{fig:path:oxford} shows that both cells produce a sequence of
action potentials. This is because the term $v_i(t - \delta_{ki})$ in
equation~(\ref{eq:path:g_k}) requires cell~$2$'s action potential to be
propagated back to cell~$1$ after a time delay. This incorrect positive
feedback behaviour is shown in Figure~\ref{fig:path:oxford} as arrows
between the action potentials.

\begin{figure}
  \centering
  
  \subfloat[Propagation from cell 1.]{
    \includegraphics[width=0.47\columnwidth]{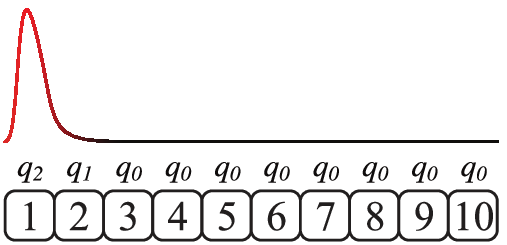}
    \label{fig:path:collision_1a}
  }
  \hfill
  \subfloat[Propagation has reached cell 5.]{
    \includegraphics[width=0.47\columnwidth]{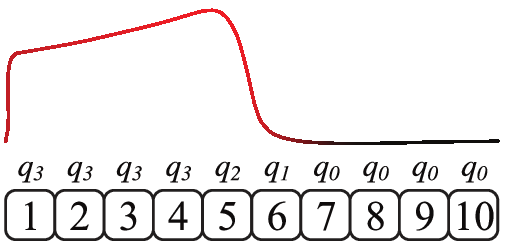}
    \label{fig:path:collision_1b}
  }
  
  \subfloat[Another propagation from cell 10.]{
    \includegraphics[width=0.47\columnwidth]{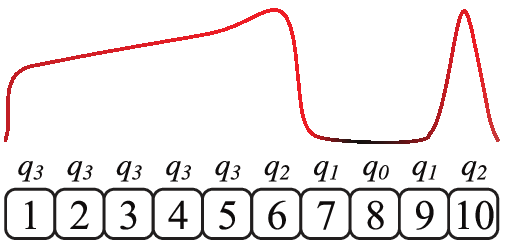}
    \label{fig:path:collision_1c}
  }
  \hfill
  \subfloat[Propagations collide.]{
    \includegraphics[width=0.47\columnwidth]{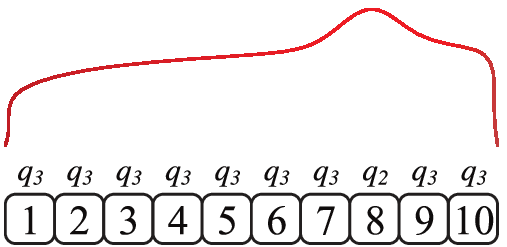}
    \label{fig:path:collision_1d}
  }

  \subfloat[Propagations annihilated.]{
    \includegraphics[width=0.47\columnwidth]{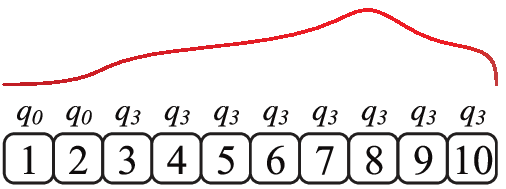}
    \label{fig:path:collision_1e}
  }
  \hfill
  \subfloat[The path model mimics the propagation along a chain of cells.]{
    \includegraphics[width=0.47\columnwidth]{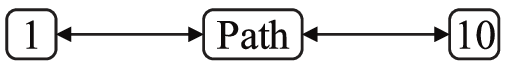}
    \label{fig:path:path}
  }

  \caption{Propagation and collision of electrical stimuli along cardiac tissue.}
  \label{fig:path:collision_1}
\end{figure}

\begin{figure*}
  \centering

  \subfloat[TA to determine which action potentials can propagate.]{
    
    \includegraphics[width=1.5\columnwidth]{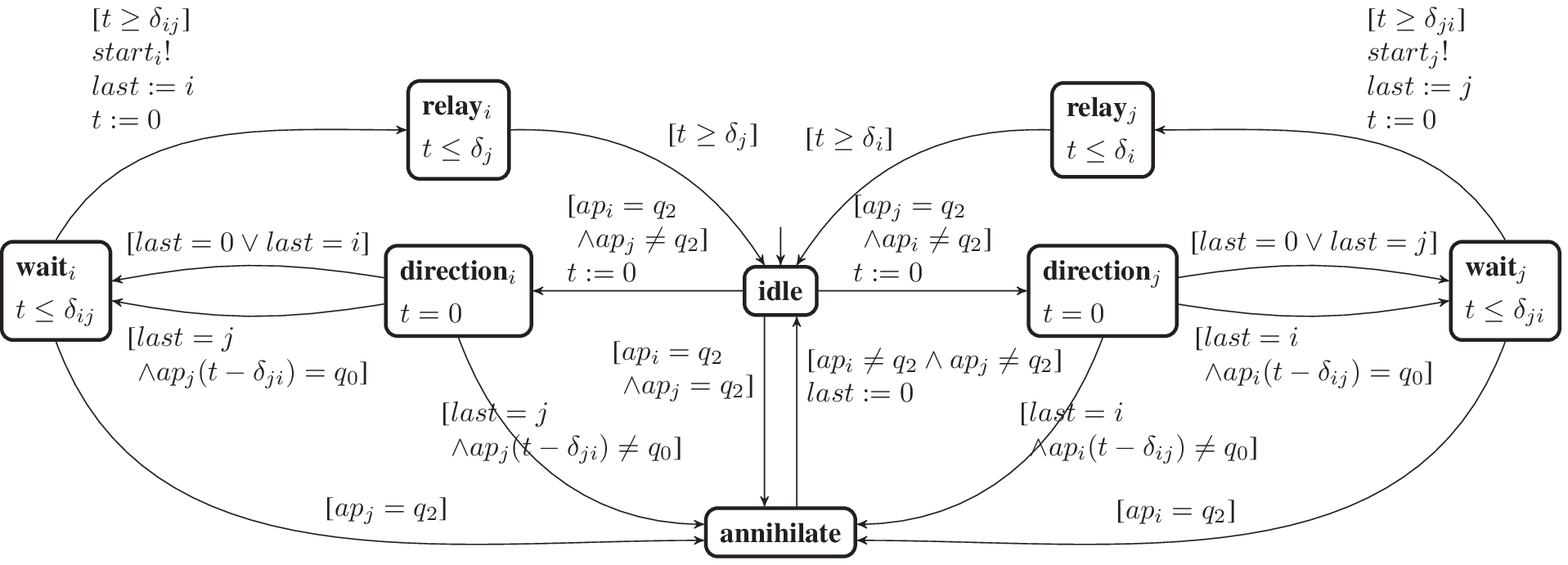}
    \label{fig:path:ta_1}
  }
  
  \subfloat[TA to relay the action potential of cell~$i$ at cell~$j$.]{

	\includegraphics[width=0.7\columnwidth]{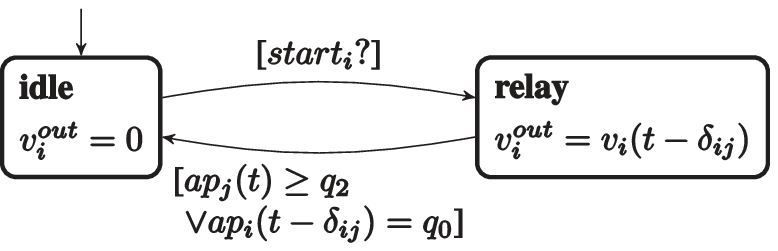}
    \label{fig:path:ta_2}
  }
  \hspace{1cm}
  \subfloat[TA to relay the action potential of cell~$j$ at cell~$i$.]{
    
    \includegraphics[width=0.7\columnwidth]{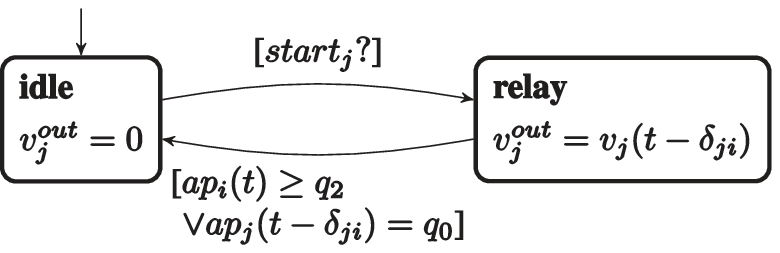}
    \label{fig:path:ta_3}
  }

  \caption{Improved path model (UoA) consisting of a TA to detect collisions and two TA to relay the action potentials.}
  \label{fig:path:ta}
\end{figure*}

To properly model the propagation of action potentials along a path, its
behaviour in real cardiac tissue needs to be
reviewed. Figure~\ref{fig:path:collision_1a} shows a chain of cardiac
cells where only cell~$1$ has entered its upstroke phase. The cells'
membrane potentials are plotted above the cells along with their
corresponding HA location.  In Figure~\ref{fig:path:collision_1a},
cell~$1$'s membrane potential starts to stimulate cell~$2$ to enter the
upstroke phase. Cell~$2$ will then stimulate its neighbour and so on.
In Figure~\ref{fig:path:collision_1b}, cell~$5$ enters the upstroke
phase, while cells~$1$ to~$4$ have entered the plateau and early
repolarisation phase. Cells~$1$ to~$4$ are unresponsive to any
electrical stimuli applied to them. This refractory feature forces an
action potential to propagate in one direction along a path. When two
action potentials propagate towards each other
(Figures~\ref{fig:path:collision_1c}), they will collide
(Figure~\ref{fig:path:collision_1d}) and annihilate each other
(Figure~\ref{fig:path:collision_1e}), i.e., the action potentials will
not pass through each other.

A computationally efficient heart model can be created by replacing
chains of cardiac cells with paths that mimic the propagation of action
potentials (Figure~\ref{fig:path:path}). The path model of
UPenn~\cite{JiangPAM14} does consider the refractory feature of cardiac
cells and can model the collision of electrical stimuli. However, only
the propagation of discrete action potential events, rather than
continuous-time signals, are modelled. Moreover, a cell can only be
stimulated by the electrical activity of one neighbour at a time. We
propose a new path model that handles the collision of continuous-time
action potentials and calculates the overall voltage induced by a cell's
neighbours with a reaction-diffusion~\cite{KleberR04} function.

\subsection{Improved Path Model}
\label{sec:path:improvement}

Our \emph{timed automaton} (TA) for determining whether an action
potential can propagate along a path is shown in
Figure~\ref{fig:path:ta_1}. To keep the path model simple, we assume
that the path is only long enough for one complete action potential to
propagate through. That is, we assume that the duration of the
propagating action potential is longer than its conduction time through
the path. An action potential from cell~$i$ will fail to propagate to
cell~$j$ under the following circumstances:
\begin{itemize}
\item Cell~$j$ enters the upstroke phase during the 
  conduction time.
  
\item A recent propagation from cell~$j$ to cell~$i$ 
  failed to stimulate cell~$i$, i.e., a partial 
  propagation, and the cells in between have not
  returned back to their resting phase.
\end{itemize}
To recognise these circumstances, the TA requires the following
information of the cardiac cells (Figure~\ref{fig:cell:improved}):
\begin{itemize}
\item $ap_i$ and $ap_j$: The phase of cell~$i$ and~$j$'s
  action potential.
  
\item $ap_i(t - \delta_{ij})$: 
  The phase of cell~$i$'s action potential when it reaches
  cell~$j$ after a conduction time of~$\delta_{ij}$.
  
\item $ap_j(t - \delta_{ji})$: 
  The phase of cell~$j$'s action potential when it reaches
  cell~$i$ after a conduction time of~$\delta_{ji}$.
\end{itemize}
In addition, the local variable \emph{last} is needed to remember the
direction of the last propagation:
\begin{equation*}
  last = 
  \left \lbrace
    \begin{array}{ll}
      0	& \text{if the last propagation was annihilated}		\\
      i	& \text{if the last propagation started from cell}~i	\\
      j	& \text{if the last propagation started from cell}~j	\\
    \end{array}
  \right .
\end{equation*}

In Figure~\ref{fig:path:ta_1}, the left half of the TA determines if an
action potential from cell~$i$ can propagate to cell~$j$. The TA begins
in the \emph{idle} location and transitions to the \emph{annihilate}
location if both cells~$i$ and~$j$ go into the upstroke phase~$q_2$.  A
transition back to the \emph{idle} location is made when both cells
exit their upstroke phase~$q_2$. If only cell~$i$ goes into the
upstroke phase~$q_2$, then a transition from the \emph{idle} to
\emph{direction}$_i$ location is made. Next, if a partial propagation
from cell~$j$ occurred recently, then a transition to the
\emph{annihilate} location is made. Otherwise, a transition to the
\emph{wait}$_i$ location is made. The TA transitions to the
\emph{annihilate} location if cell~$j$ enters the upstroke phase~$q_2$
during cell~$i$'s conduction time~$\delta_{ij}$.  Otherwise, the TA
remains in the \emph{wait}$_i$ location until the conduction time has
elapsed, at which point the signal $start_i$ is emitted (denoted by 
``$!$'') and a transition
to the \emph{relay}$_i$ location is made. Next, a transition to the
\emph{idle} location is made after a time of $\delta_j$ has elapsed to
ignore cell~$j$'s upstroke phase. The right half of the TA uses similar
logic to determine if an action potential from cell~$j$ can propagate to
cell~$i$.

The TA in Figure~\ref{fig:path:ta_2} relays cell~$i$'s action potential at
cell~$j$ when the signal $start_i$ is received (denoted by ``$?$''). 
The relay stops when cell~$j$ has successfully depolarised or when 
cell~$i$'s action potential reaches the
resting phase~$q_0$. Similar logic is used by the TA in
Figure~\ref{fig:path:ta_3} to relay cell~$j$'s action potential at
cell~$i$ when the signal $start_i$ is emitted.  A cell can be connected
to multiple neighbours and, therefore, receive multiple action
potentials simultaneously.  We determine the voltage induced at cell~$k$
by its~$n$ neighbours with the following
reaction-diffusion~\cite{KleberR04} function:
\begin{equation}
	\label{eq:path:h_k}
	h_k(\vec{v}) = \sum^n_{i = 1} \frac{\Gamma_{ik} \sigma_{ik}}{A_m C_m}(v^{out}_i - v_k)
\end{equation}

\noindent
where~$\Gamma_{ik}$ is the cross-sectional area (units of $mm^2$) from
cell~$i$ to~$k$, $\sigma_{ik}$ is the electrical conductivity (units of
$mS/mm$) from cell~$i$ to~$k$, $A_m$ is cell~$k$'s surface area to
volume (units of $mm^{-2}$), $C_m$ is cell~$k$'s specific membrane
capacitance (units of $\mu F/mm^2$), and $v^{out}_i$ is cell~$i$'s
membrane potential after propagating along the path.

\subsection{Discussion}
We have developed an improved path model specifically for
continuous-time action potentials. We can successfully model
propagations that are annihilated by a full or partial propagation from
the opposite direction. This allows us to model conduction block that
only occur in one direction, e.g., due to the source-sink
relationship~\cite{Spector13}, which the UPenn path
model~\cite{JiangPAM14} can not. Bartocci et al.~\cite{BartocciCDESG09}
modelled cardiac tissue with a high-resolution grid of Stony Brook
cells~\cite{YeESG08}. In that model, the cells are connected
electrically by a reaction-diffusion equation because the conduction
times between the cells are not significant.

\begin{figure}
	\centering

	\subfloat[Full and partial propagations.]{
		\includegraphics[width=0.15\columnwidth]{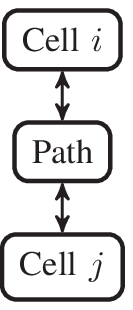}
		\includegraphics[width=0.8\columnwidth]{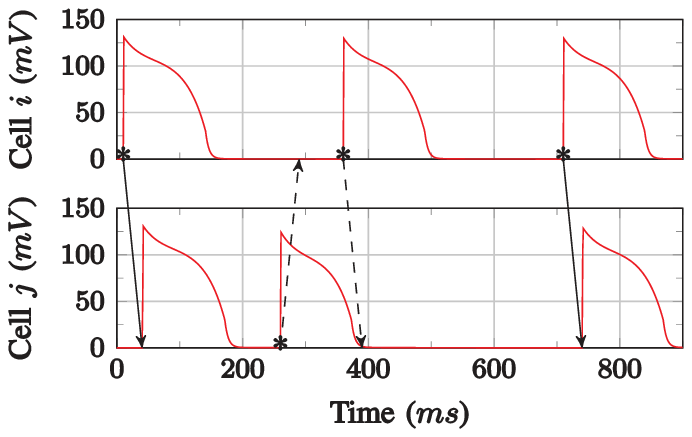}
		\label{fig:path:example_1}
	}

	\subfloat[Multiple neighbours needed to stimulate a cell.]{
		\includegraphics[width=0.35\columnwidth]{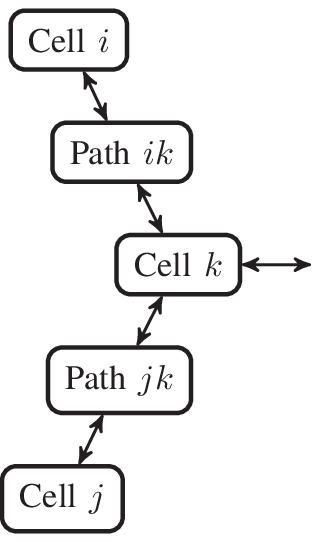}
		\includegraphics[width=0.60\columnwidth]{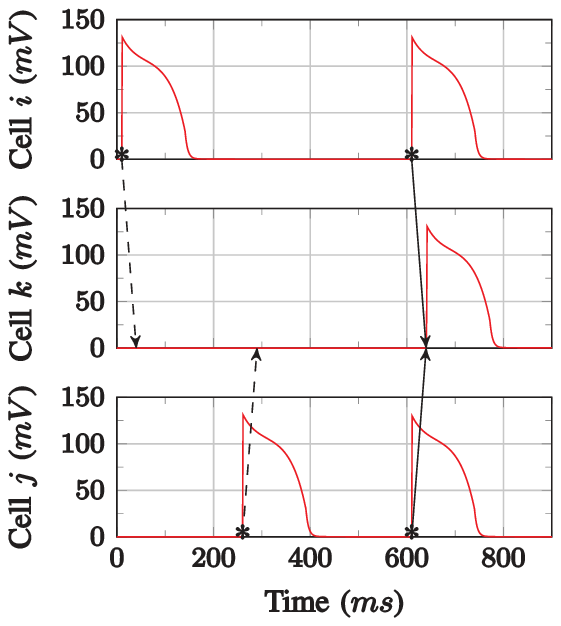}
		\label{fig:path:example_2}
	}

	\caption{Examples of propagation behaviour that can be modelled by
			 the improved path model, but not by the Oxford~\cite{ChenDKM14} or
			 UPenn~\cite{JiangPAM14} path models.}
	\label{fig:path:example}
\end{figure}

Figure~\ref{fig:path:example_1} demonstrates two cells~$i$ and~$j$ with
a conduction block from cell~$j$ to~$i$. The conduction time between the
cells is $30ms$. At $10ms$, cell~$i$ generates an action potential that
successfully stimulates cell~$j$. At $260ms$, cell~$j$ generates an
action potential that does not stimulate cell~$i$ because of conduction
block. Shortly after at $360ms$, cell~$i$ generates an action potential
that does not reach cell~$j$ because the cells along the path have not
returned back to the resting phase~$q_0$.  At $710ms$, cell~$i$
generates an action potential that successfully stimulates cell~$j$.

Figure~\ref{fig:path:example_2} demonstrates a cell~$k$ that will only
stimulate if the action potentials of its neighbours, cells~$i$ and~$j$, arrive
together. At times $10ms$ and $260ms$, cells~$i$ and~$j$, respectively, 
generate action potentials that take $30ms$ to propagate to cell~$k$. 
However, the voltage induced on cell~$k$ is neither strong enough nor 
long enough to stimulate cell~$k$. At 
time $610ms$, both neighbours generate action potentials that arrive
simultaneously at cell~$k$. This time, the voltage induced on cell~$k$ 
is enough to stimulate it.

\begin{figure*}
	\centering

	\hfill
	\subfloat[Schematic of the fast and \newline slow pathways around the AV node.]{
		\includegraphics[width=0.40\columnwidth]{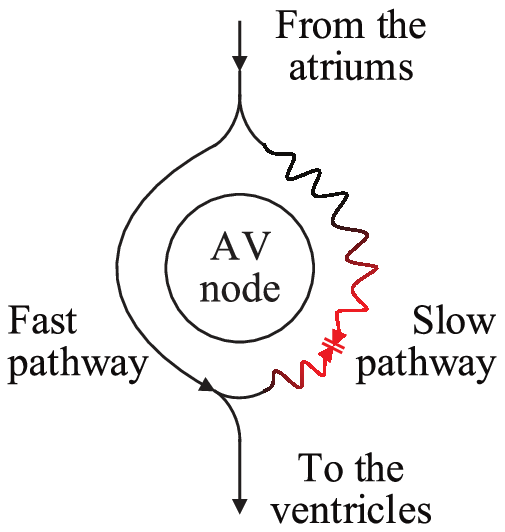}
		\label{fig:path:re_entry_1a}
	}
	\hspace{2.3cm}
	\subfloat[Re-entry due to an early stimulus from the atriums.]{
		\includegraphics[width=0.40\columnwidth]{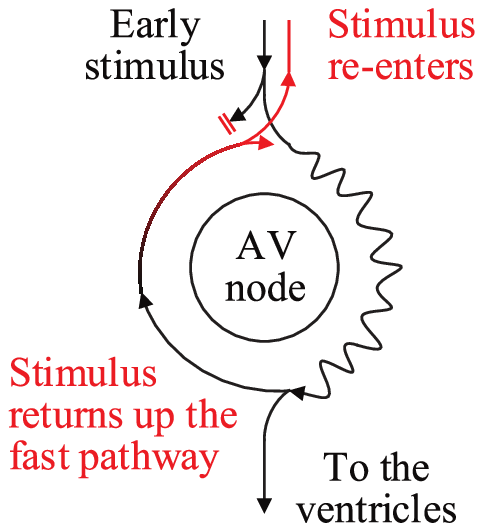}
		\label{fig:path:re_entry_1b}
	}
	\hspace{0.8cm}

	\subfloat[Normal behaviour simulated. Cell 1 stimulated at $10$ and $260ms$.]{
		\includegraphics[width=0.4\columnwidth]{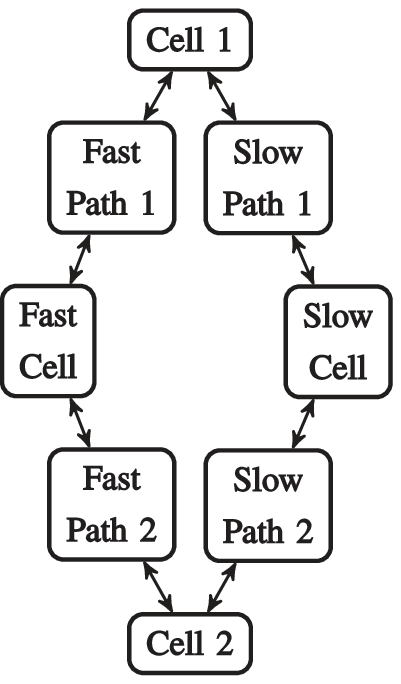}
		\includegraphics[width=0.75\columnwidth]{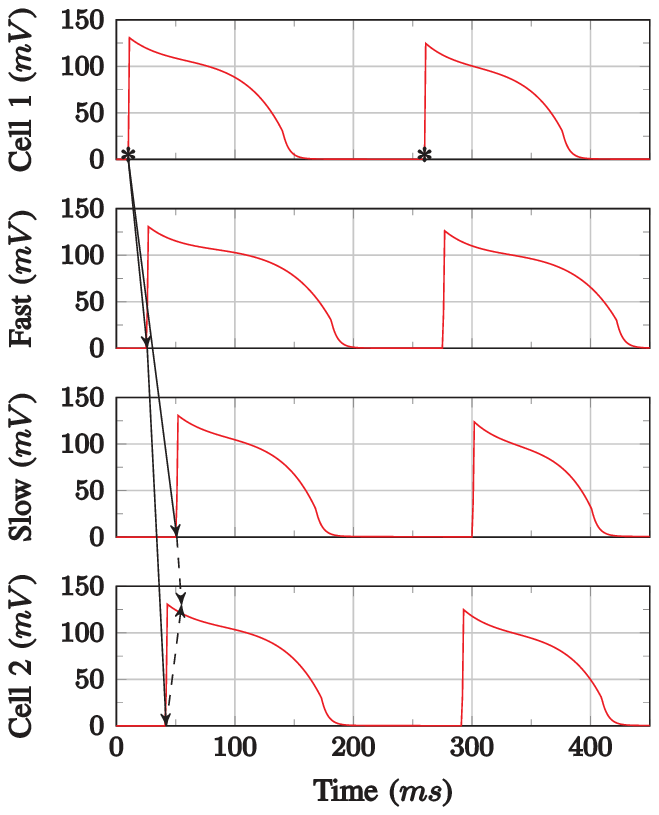}
		\label{fig:path:re_entry_2}
	}
	\hfill
	\subfloat[Abormal behaviour simulated. Cell 1 stimulated at $10$ and $160ms$.]{
		\includegraphics[width=0.75\columnwidth]{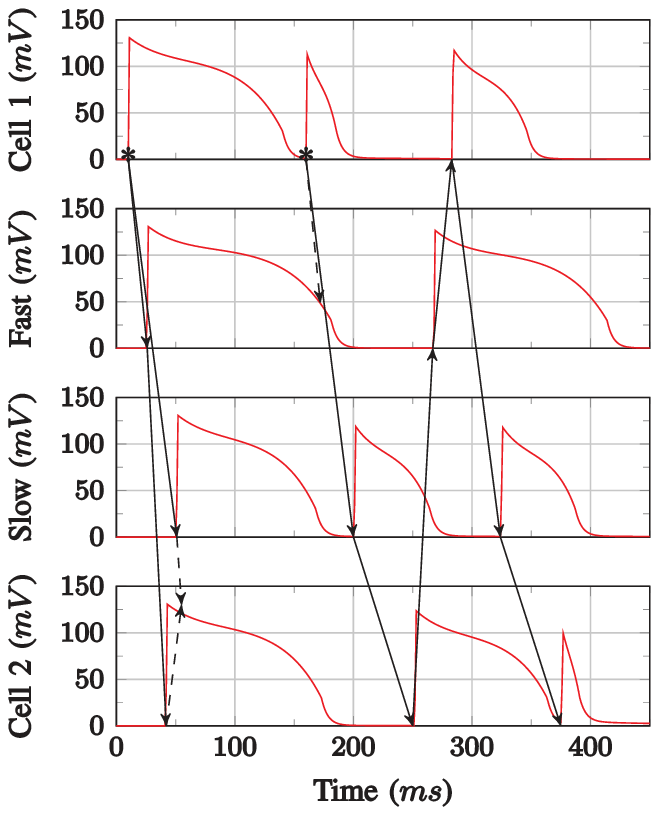}
		\label{fig:path:re_entry_3}
	}
  
	\caption{Atrioventricular node re-entrant tachycardia.}
	\label{fig:path:re_entry}
\end{figure*}

Figure~\ref{fig:path:re_entry} demonstrates atrioventricular node
re-entrant tachycardia (AVNRT~\cite{ManiP14}), caused by dual pathways
around the atrioventricular (AV) node shown in
Figure~\ref{fig:path:re_entry_1a}. An electrical stimulus propagates
faster down the left pathway than the right pathway, called the
\emph{fast} and \emph{slow} pathways, respectively. In normal cardiac
behaviour, shown in Figure~\ref{fig:path:re_entry_1a}, an electrical
stimulus from the atriums splits and propagates down the fast and slow
pathways. The stimulus in the fast pathway will reach the bottom of the
AV node before the stimulus in the slow pathway. The stimulus in the
fast pathway continues to propagate down into the ventricles and up the
slow pathway where it will annihilate the existing stimulus. Thus, only
the stimulus that propagates down the fast pathway will enter the
ventricles. Figure~\ref{fig:path:re_entry_2} demonstrates our ability to
replicate this normal behaviour by modelling the dual pathways around
the AV node with only four cells and paths.

An early electrical stimulus from the atriums can end up propagating
continuously around the AV node, shown in
Figure~\ref{fig:path:re_entry_1b}. This is because the early stimulus
cannot propagate down the fast pathway, which has cells that are still
repolarising. The cells in the fast pathway take longer to repolarise
than those in the slow pathway and, hence, have longer action potential
durations. Due to this, the stimulus can propagate down the slow pathway
because its cells have repolarised. By the time the stimulus reaches the
bottom of the AV node, the cells in the fast pathway have reached their
resting phase. Thus, the stimulus continues by propagating up the fast
pathway, during which time the cells in the slow pathway reach their
resting phase. The stimulus continues by propagating back down the slow
pathway and the cycle continues. Every time the stimulus passes the
bottom and top of the AV node, a stimulus is propagated into the
ventricles and atriums, respectively. Figure~\ref{fig:path:re_entry_3}
demonstrates our ability to replicate this abnormal behaviour 
by modelling the dual pathways around the AV node with only
four cells and paths.


\section{UoA Heart Model}
\label{sec:heart}

Following the approach of Oxford's heart model~\cite{ChenDKM14}, we
model the cardiac conduction system with a two-dimensional network of
$33$ cardiac cells, shown in Figure~\ref{fig:background:heart}. Our
heart model uses the improved cardiac cell and path models
(Sections~\ref{sec:cell:improvement_1}, \ref{sec:cell:improvement_2},
and \ref{sec:path:improvement}). The following cell and path model 
parameters were modified to capture the range of action potential
durations, conduction velocities, and electrical conductivities 
present in the conduction system:
\begin{itemize}
	\item The value of $\alpha^3_y$ in Figure~\ref{fig:cell:improved}
		  was increased to increase the cell's rate of repolarisation 
		  and, therefore, decrease its action potential duration.
	\item The values of $\delta_{ij}$ and $\delta_{ji}$ in Figure~\ref{fig:path:ta}
		  were increased to increase the conduction time and, therefore,
		  decrease the conduction velocity.
	\item The value of $\sigma_{ik}$ in equation~(\ref{eq:path:h_k}) 
		  was increased to increase the conductivity.
\end{itemize}
Cardiac electrophysiology data in the literature~\cite{DurrerDFJMA70,SiggIXH10} 
was used to estimate the parameters. The data of interest include the
morphology of myocyte action potentials and the conduction velocities of
electrical stimuli along the pathways. Table~\ref{tab:heart:uoa_oxford}
provides a qualitative comparison of Oxford's heart
model~\cite{ChenDKM14} with our improved heart model (UoA). Thanks to
improvements in our cell and path models, we are able to model more
heart conditions and with realistic results. This is
demonstrated in the next section where we provide simulation results of
the common arrhythmias described in Section~\ref{sec:background:arrhythmia}.

\begin{table}
  \centering
  \renewcommand{\arraystretch}{1.3}
  \caption{Detailed qualitative comparison of heart models from UoA and Oxford.}
  \label{tab:heart:uoa_oxford}
  
  \begin{tabular}{| L{1.5cm} | C{3cm} | C{3cm} |}
    \cline{2-3}
    \multicolumn{1}{L{1.5cm}|}{}													& {\bf UoA} 														& {\bf Oxford~\cite{ChenDKM14}} \\
    \hline
    {\bf Heart Anatomy} (Figure~\ref{fig:background:heart}) 						& \multicolumn{2}{c|}{2D conduction system} 	 											\\ \hline
    {\bf Morphology of action potentials} (Figure~\ref{fig:cell:ap_comparison})		& Realistic because a more recent Stony Brook HA model~\cite{YeESG08} is used (with the improvements from Sections~\ref{sec:cell:improvement_1} and~\ref{sec:cell:improvement_2})	& Unrealistic because a simplified Stony Brook HA model~\cite{YeEGS05} is used		\\ \hline
    {\bf Electrical Restitution} (Figure~\ref{fig:cell:restitution})				& Shows good dynamic behaviour 													& Shows limited dynamic behaviour 	\\ \hline
    \multirow{2}{1.5cm}{\newline {\bf Path Model} (Figure~\ref{fig:path:example})}	& \multicolumn{2}{C{6.5cm}|}{Electrical propagation is time delayed, bi-directional, and anisotropic.} 												\\ \cline{2-3}
                                                                                                                                        & Correctly disallows colliding electrical stimuli from passing through each other 															& Incorrectly allows colliding electrical stimuli to pass through each other		\\ \hline
    {\bf Disease Conditions} (Figures~\ref{fig:results:block} to~\ref{fig:results:tachycardia})						& Bradycardia and tachycardia by modifying the SA rate, AV node re-entrant tachycardia, VA conduction, heart blocks, and long Q-T syndrome											& Bradycardia and tachycardia by modifying the SA rate, and heart blocks 	\\
    \hline
  \end{tabular}
\end{table}


\section{Simulation Results}
\label{sec:results}

\begin{figure*}

	\centering
	\includegraphics[width=1.8\columnwidth]{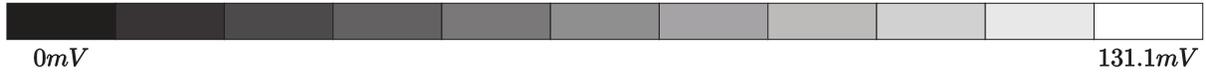}

	\caption{Grayscale colourmap used to represent the voltage of each node.}
	\label{fig:results:grayscale}
\end{figure*}

\begin{figure*}
	\centering

	\input{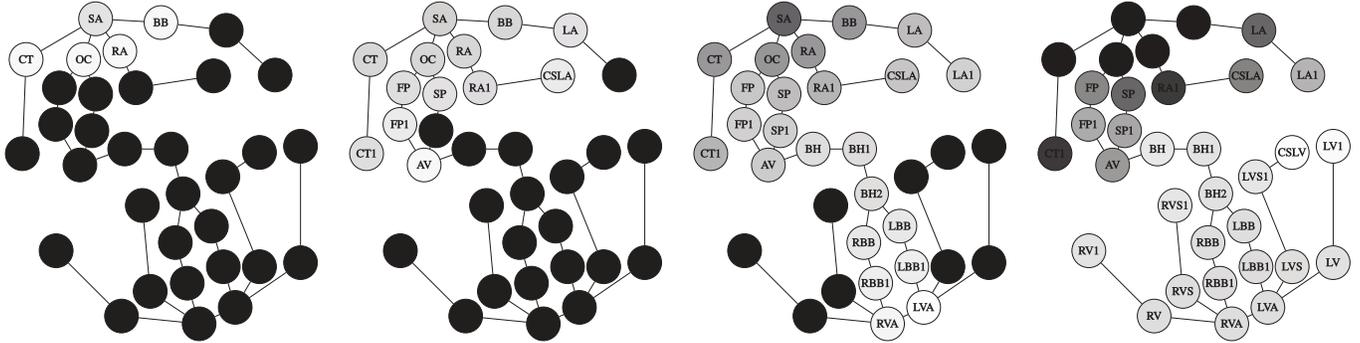}

	\caption{Simulation of a normal cardiac cycle. The SA node depolarises at $10ms$.}
	\label{fig:results:normal}
\end{figure*}

We implemented our heart model in
MathWorks\textsuperscript{\textregistered}
Simulink\textsuperscript{\textregistered} and
Stateflow\textsuperscript{\textregistered} R2015b Pro. 
A fixed simulation time step of $0.0005ms$ and the
Dormand-Prince ODE45 solver was used. 
Figure~\ref{fig:results:normal} shows the propagation of
electrical stimuli through the cardiac conduction system 
(Figure~\ref{fig:background:heart}) during a
normal cardiac cycle. The membrane potential of each 
node is denoted by a shade of grey, according to 
Figure~\ref{fig:results:grayscale}. In the following, 
we provide simulation results of the arrhythmias 
described in Section~\ref{sec:background:arrhythmia}.

\begin{figure*}
	\centering

	\input{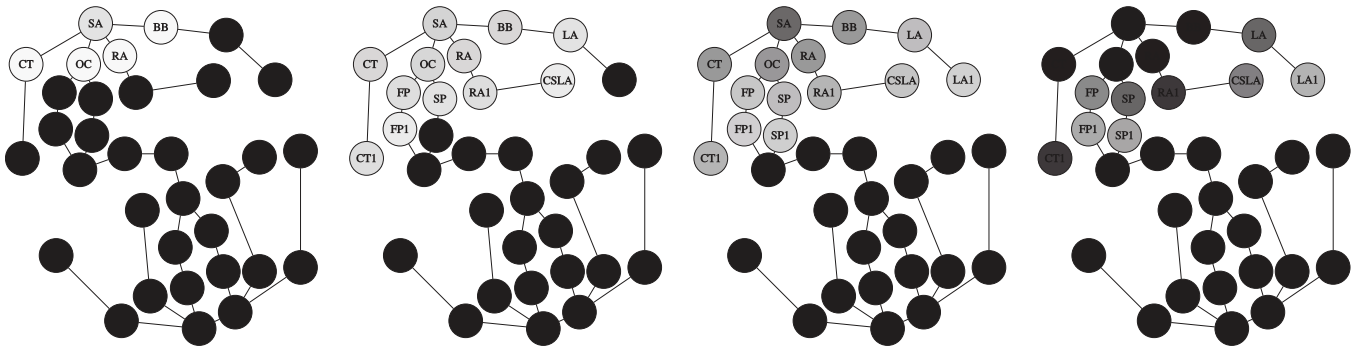}

	\caption{Simulation of heart block. The SA node depolarises at $10ms$.}
	\label{fig:results:block}
\end{figure*}

\begin{figure*}
	\centering

	\input{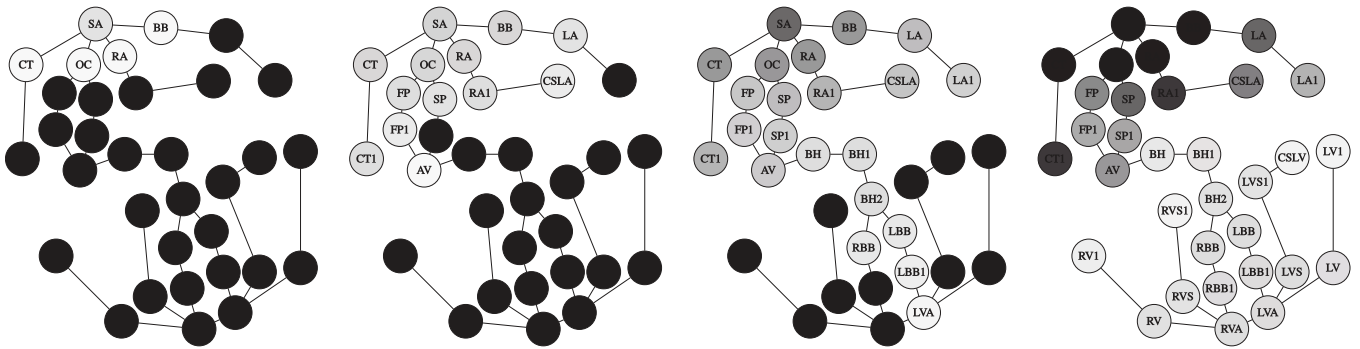}

	\caption{Simulation of right bundle branch block. The SA node depolarises at $10ms$.}
	\label{fig:results:branch_block}
\end{figure*}

\begin{figure*}
	\centering

	\input{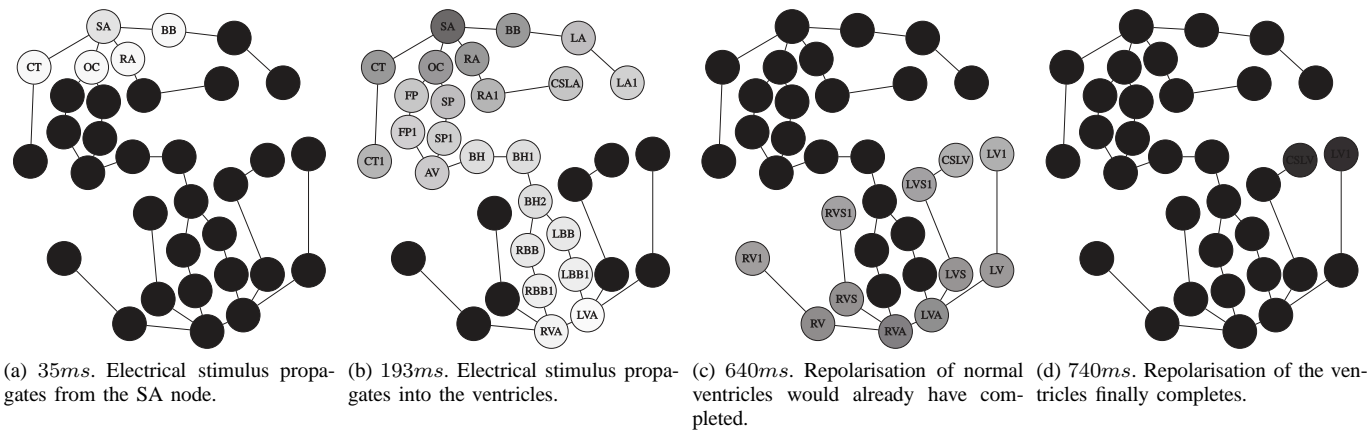}

	\caption{Simulation of long Q-T syndrome. The SA node depolarises at $10ms$.}
	\label{fig:results:qt_syndrome}
\end{figure*}

\begin{figure*}
	\centering

	\input{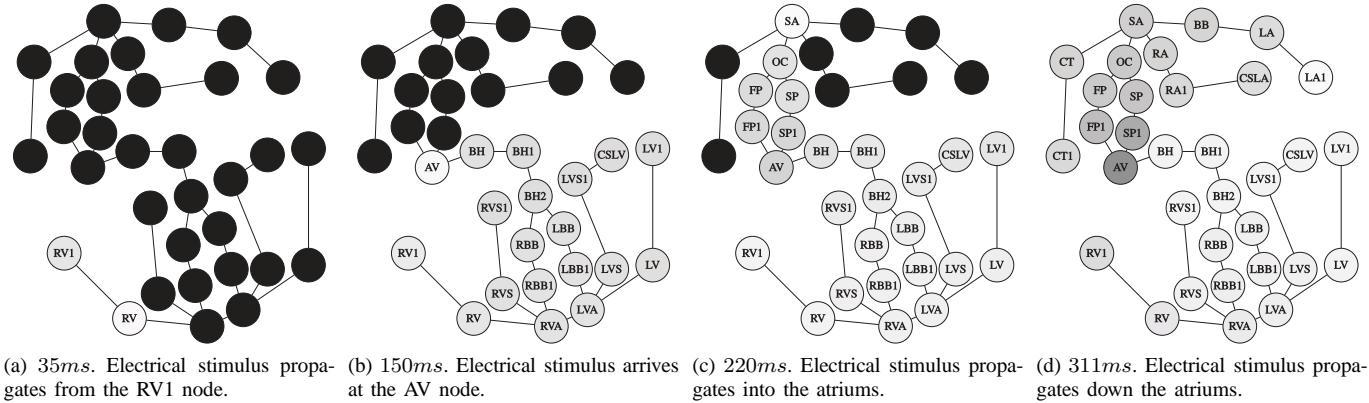}

	\caption{Simulation of VA conduction. The RV1 node depolarises at $10ms$.}
	\label{fig:results:va_conduction}
\end{figure*}

\begin{figure*}
	\centering

	\input{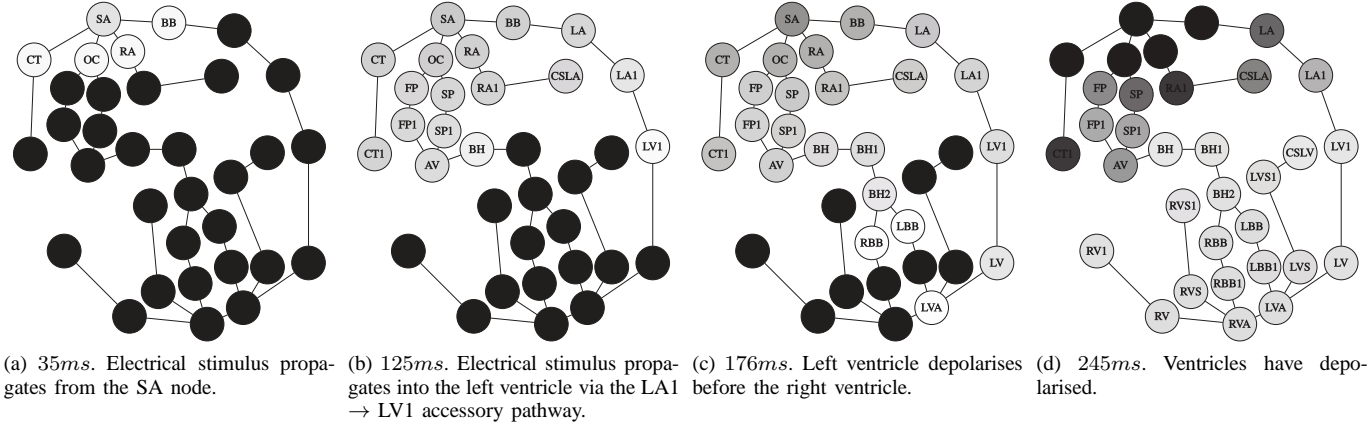}

	\caption{Simulation of Wolff-Parkinson-White syndrome. The SA node depolarises at $10ms$.}
	\label{fig:results:wpw_syndrome}
\end{figure*}

\begin{figure*}
	\centering

	\input{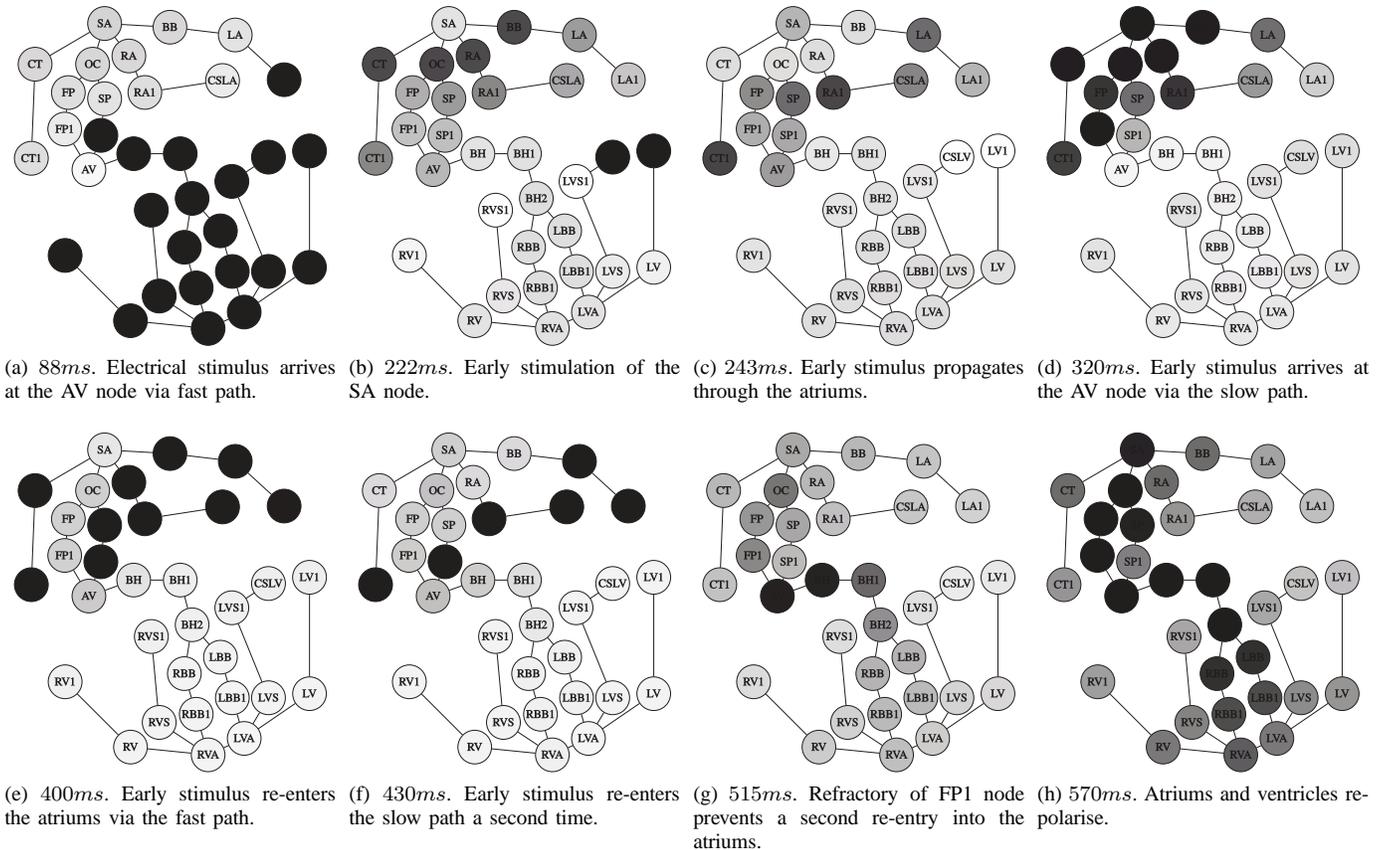}

	\caption{Simulation of AV node re-entrant tachycardia. The SA node depolarises at $10$ and $220ms$.}
	\label{fig:results:tachycardia}
\end{figure*}

\subsection{Heart Block}
This is modelled by reducing the voltage that the
neighbouring atrial cells induce on the AV node and by increasing the
conduction time through the AV
node. Figure~\ref{fig:results:block} shows a simulation of heart block
where an electrical stimulus does not propagate through the AV node.

\subsection{Bundle Branch Block}
This is modelled similarly to heart block. For one of the
bundle branches, the voltage induced by the
AV node is reduced and the conduction time
along the branch is increased. Figure~\ref{fig:results:branch_block} 
shows a simulation of bundle branch block where propagation 
down the right branch is slower than the left branch.

\subsection{Long Q-T Syndrome}
This is modelled by decreasing the value of $\alpha^3_y$ in 
the ventricular cells to decrease their rate of repolarisation,
thereby creating a longer action potential duration.
Figure~\ref{fig:results:qt_syndrome} shows a simulation of the 
syndrome.

\subsection{VA Conduction}
This is modelled by stimulating a ventricular node before
the SA node generates a stimulus. Figure~\ref{fig:results:va_conduction}
shows a simulation of VA conduction by stimulating the RV1 
node in the right ventricle.

\subsection{Wolff-Parkinson-White Syndrome}
This is modelled by creating an additional pathway between
the left atrium and left ventricle. Figure~\ref{fig:results:wpw_syndrome}
shows a simulation of the syndrome, which causes the left 
ventricle to depolarise earlier than usual.

\subsection{AV Node Re-entrant Tachycardia}
This is modelled by the fast and slow pathways around the SA 
node, demonstrated previously in Figure~\ref{fig:path:re_entry}. 
Figure~\ref{fig:results:tachycardia} simulates re-entrant 
tachycardia by stimulating the SA node a second time with an 
early stimulus. 

\subsection{Time to Simulate the Cell and Heart Models}
The execution time needed to simulate the cell and heart 
models of Oxford and UoA are compared in this section.
For a fair comparison, we wanted to avoid the execution 
overhead of the Simulink\textsuperscript{\textregistered} 
environment. Thus, for each model, Simulink Coder\textsuperscript{TM}
was used to generate a single-threaded executable C program 
that is optimised for execution speed. Each C program 
was then executed on a single core of a $3.40GHz$ Intel 
Core i7-4770 processor with $16GB$ of RAM and running 
Microsoft Windows 7 Enterprise. Each reported execution 
time is the average of ten runs of the same program.

The Oxford cell model takes $12.376s$ to simulate $60s$ of cell
activity with a $1s$ base cycle length, compared with $9.586s$ for
the UoA improved cell model. The Oxford cell model takes
longer to simulate because it uses a Stateflow\textsuperscript{\textregistered}
block to manually check the transiton guards and produce a
corresponding Boolean event whenever a guard is true. The
Boolean event triggers another Stateflow\textsuperscript{\textregistered} 
block to take transitions between the hybrid automaton states. 
For our UoA improved cell model, the hybrid automaton is modelled 
by a single Stateflow\textsuperscript{\textregistered} block. 
The hybrid automaton transition guards are modelled directly as 
Stateflow\textsuperscript{\textregistered} transition
conditions. Stateflow\textsuperscript{\textregistered} 
(automatically) handles the checking of transitions, which 
leads to a more efficient simulation, even though the Oxford 
cell model only uses the $v_x$ variable of the Stony Brook 
2005 cell model. Moreover, the UoA improved cell model uses 
Simulink\textsuperscript{\textregistered} components that are 
more efficient to simulate wherever possible.

The Oxford and UoA heart models use the same arrangement of 
$33$ nodes, as shown in Figure~\ref{fig:background:heart}.
The Oxford heart model has a total of $34$ paths between the
nodes and takes $8.516s$ to simulate $1s$ of a normal
cardiac cycle. The UoA improved heart model
has a total of $68$ forward and backward paths between the 
nodes and takes $30.771s$ to simulate $1s$ of a normal
cardiac cycle and $31.295s$ to simulate $1s$ of AV node
re-entrant tachycardia. The UoA improved heart model takes longer to
simulate because it implements a much more complex path
model that considers the backward propagation and
annihilation of electrical impulses throughout the cardiac
conduction system. Thus, it is imperative to further improve
the efficiency of the path model without sacrificing its 
predictive power. The execution time of the UoA improved
heart model could be reduced by parallelising the simulation
of the nodes over multi-cores~\cite{BartocciCGGSF11}.

\section{Conclusion}
\label{sec:conclusion}

We carefully reviewed the heart models designed for the closed-loop
testing of pacemakers and identified key areas of improvement. To this
end, we advanced the state-of-the-art in three ways: (1) stabilising and
enforcing a minimum action potential duration on the Stony Brook cardiac
cell model, (2) developing a path model that handles the partial and
full propagation of continuous-time action potentials, and (3)
demonstrating the predictive power of our heart model through the
extensive simulation of arrhythmias. We are able to model many more
arrhythmias than are possible with the existing heart models.

For future work, we look to implement our heart model with more
computationally efficient methods to achieve the goal of real-time heart
simulation, i.e., heart emulation. We also look to investigate the
automatic generation and parameterisation of a network of cardiac cells
and paths for patient-specific heart modelling.





\section*{Acknowledgment}
This research was funded by the University of Auckland FRDF
Postdoctoral grant No. $3707500$. The authors would like to
express their gratitude to Prof. Marta Kwiatkowsk and her
research group from the University of Oxford for providing
an implementation of their heart model.

\ifCLASSOPTIONcaptionsoff
  \newpage
\fi

\begin{IEEEbiography}[{\includegraphics[width=2in,height=1.25in,clip,keepaspectratio]{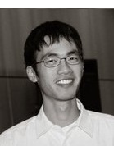}}]{Eugene Yip}
	received his B.E. (Hons) and Ph.D degrees in
	electrical and computer systems engineering from the
	University of Auckland, New Zealand. At the start of
	2015, he joined the Heart-on-FPGA research group in the
	Department of Electrical and Computer Engineering,
	University of Auckland, as a research assistant working
	on the real-time modelling of cardiac electrical
	activity. In the middle of 2015, he joined the SWT
	research group at the University of Bamberg as a
	research assistant working on synchronous languages. His
	current research interests include synchronous
	mixed-criticality systems, parallel programming, static
	timing analysis, formal methods for organ modelling, and
	biomedical devices.
\end{IEEEbiography}

\begin{IEEEbiography}[{\includegraphics[width=1in,height=1.25in,clip,keepaspectratio]{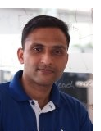}}]{Sidharta Andalam}
	received his Ph.D degree from the University of
	Auckland, New Zealand, in 2013, where his thesis focused
	on developing a predictable platform for safety-critical
	systems. He is currently a research fellow in embedded
	systems at the University of Auckland, New Zealand. His
	principle research interest is in the design,
	implementation, and analysis of safety-critical
	applications. He has worked at TUM CREATE, Singapore,
	exploring safety-critical applications in the automotive
	domain.
\end{IEEEbiography}

\begin{IEEEbiography}[{\includegraphics[width=1in,height=1.25in,clip,keepaspectratio]{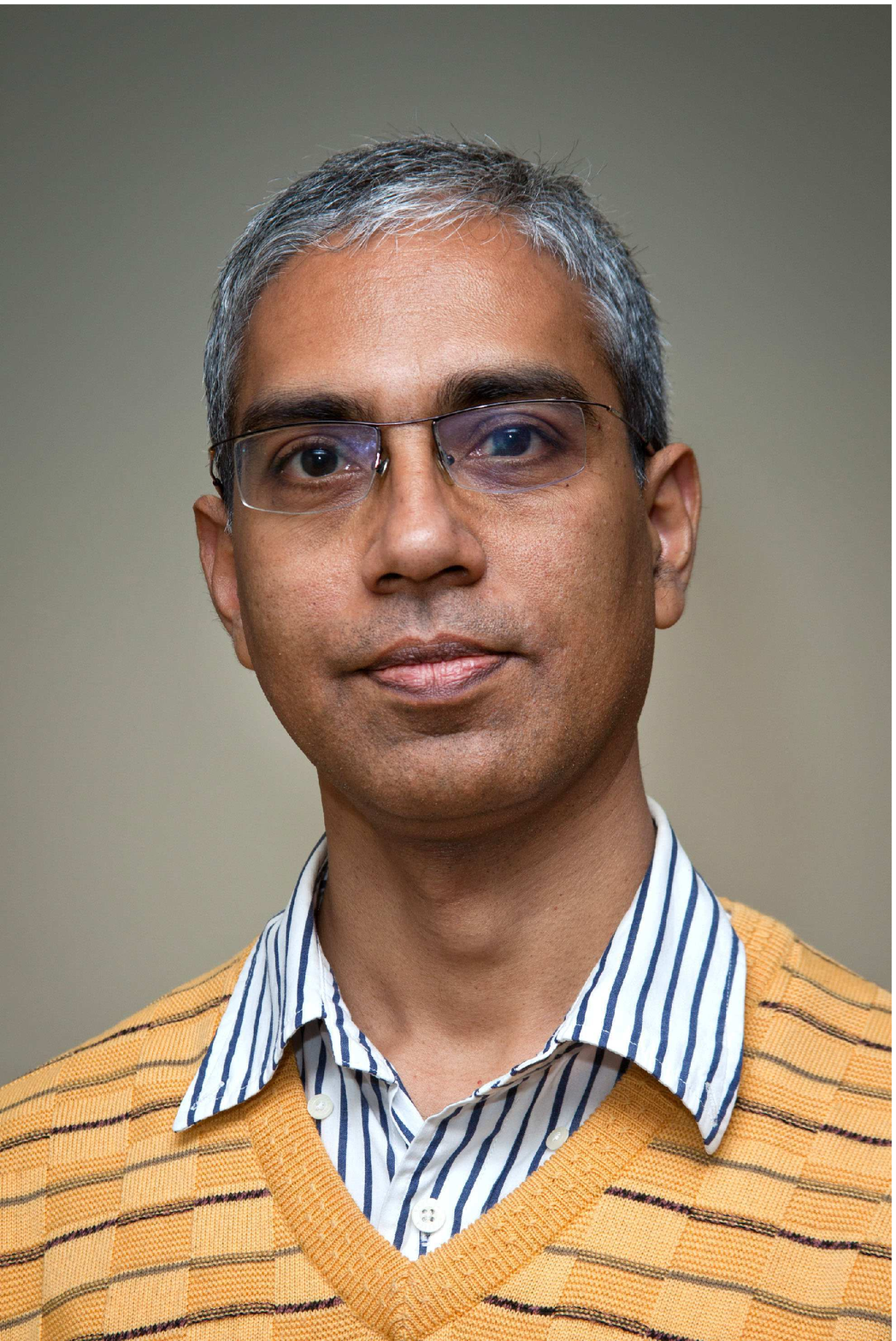}}]{Partha S. Roop}
	received his Ph.D degree in computer science (software
	engineering) from the University of New South Wales,
	Sydney, Australia, in 2001. He is currently an Associate
	Professor and is the Director of the Computer Systems
	Engineering Program with the Department of Electrical
	and Computer Engineering, the University of Auckland, New
	Zealand. 
	Partha is an associated team member of the SPADES team
	INRIA, Rhone-Alpes, France, and held a visiting position in
	CAU Kiel, Germany, and Iowa State University, USA.
	His research interests include the design and
	verification of embedded systems. In particular, he is
	developing techniques for the design of embedded
	applications in automotive, robotics, and intelligent
	transportation systems that meet functional-safety
	standards.
\end{IEEEbiography}

\begin{IEEEbiography}[{\includegraphics[width=1in,height=1.25in,clip,keepaspectratio]{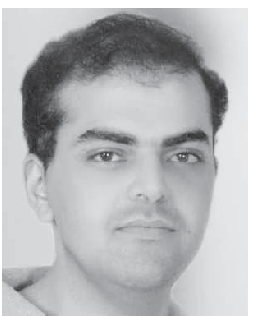}}]{Avinash Malik}
	is a lecturer at the University of Auckland, New
	Zealand. His main research interest lies in programming
	languages for multicore and distributed systems and
	their formal semantics and compilation. He has worked at
	organisations such as INRIA in France, Trinity College
	Dublin, IBM research Ireland, and IBM Watson on the design
	and the compilation of programming languages. He holds B.E.
	and Ph.D degrees from the University of Auckland.
\end{IEEEbiography}

\begin{IEEEbiography}[{\includegraphics[width=1in,height=1.25in,clip,keepaspectratio]{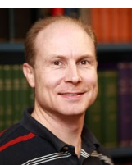}}]{Mark Trew}
	attained a Bachelor of Engineering in Engineering
	Science in 1992 and a Ph.D Engineering Science in 1999,
	both from the University of Auckland, New Zealand. He is
	currently a Senior Research Fellow at the Auckland
	Bioengineering Institute. Mark constructs computer
	models and analysis tools for interpreting and
	understanding detailed images of cardiac tissue and
	cardiac electrical activity.
\end{IEEEbiography}

\begin{IEEEbiography}[{\includegraphics[width=1in,height=1.25in,clip,keepaspectratio]{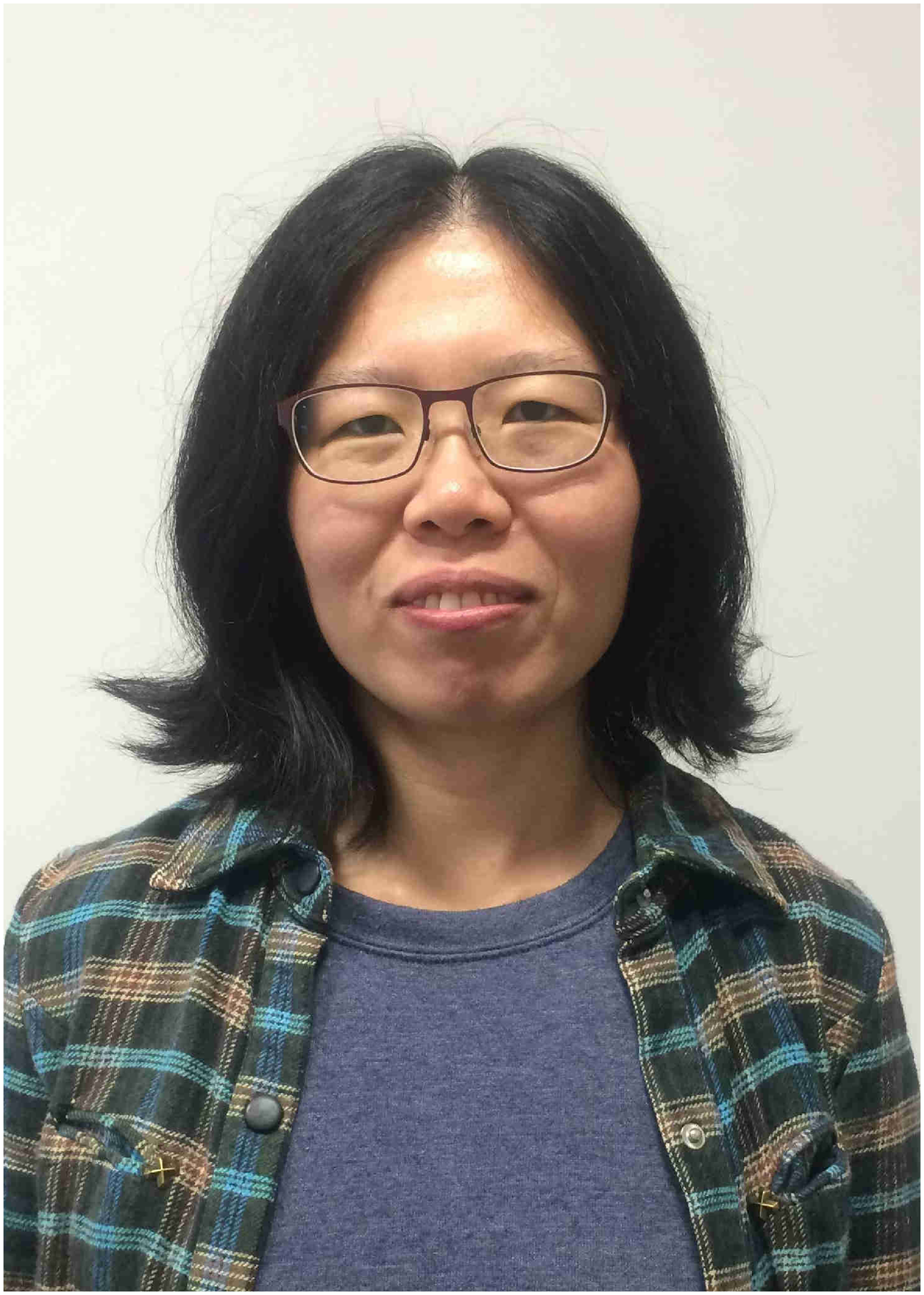}}]{Weiwei Ai}
	received the B.S. degree in 2003 from Qingdao University, 
	China, and the M.E. degree in 2006 from Beijing University 
	of Technology, China. From 2006 to 2013, she worked as a 
	reliability and failure analysis engineer in CEC Huada 
	Electronic Design Co.,Ltd. She is currently working towards 
	the Ph.D. degree at the the University of Auckland, New 
	Zealand. Her research interests are in verification and 
	validation with a focus on medical devices.   
\end{IEEEbiography}

\begin{IEEEbiography}[{\includegraphics[width=1in,height=1.25in,clip,keepaspectratio]{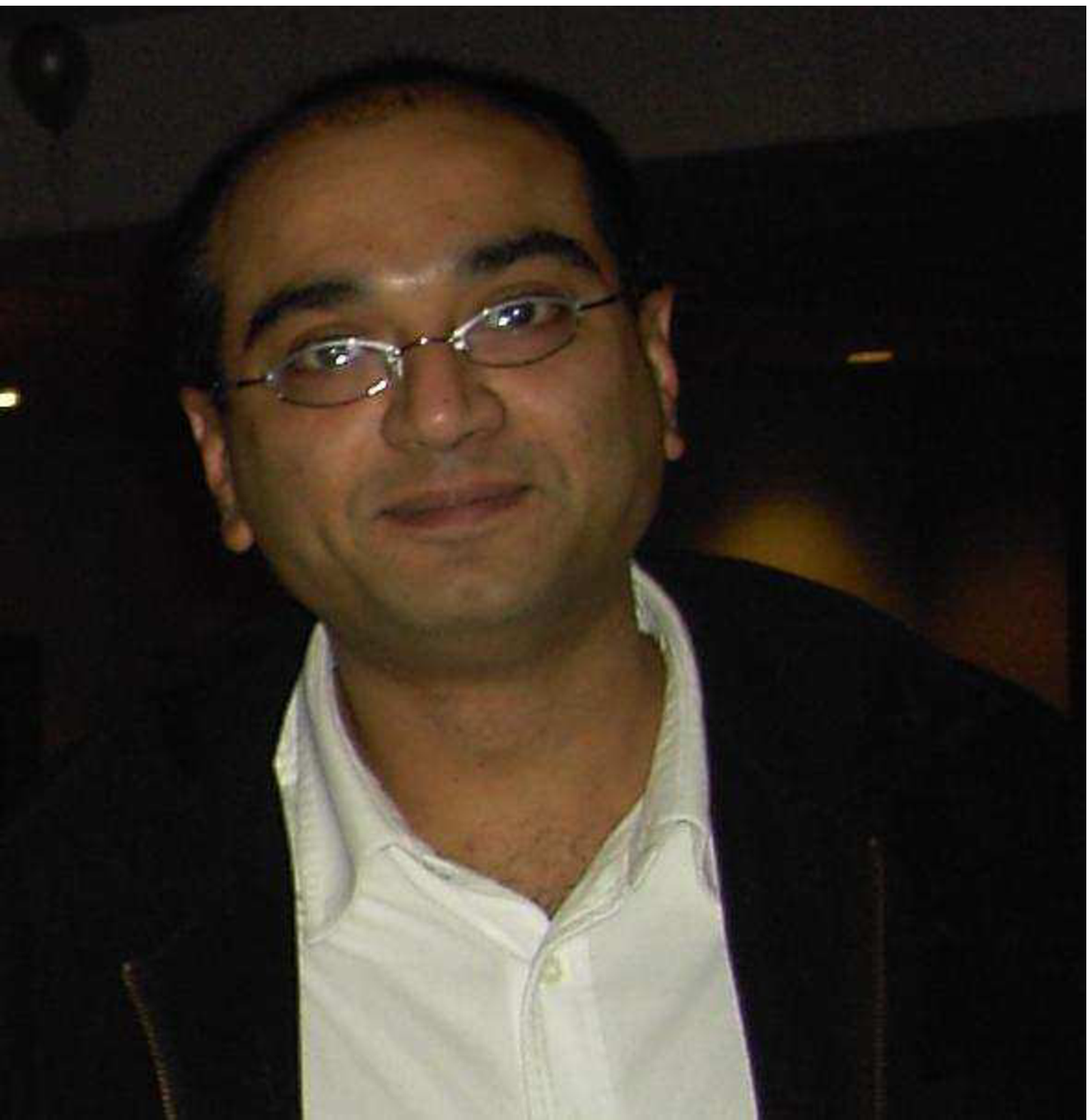}}]{Nitish Patel}
	received the B.E. degree from Mangalore University, 
	Karnataka, India, and the Ph.D. degree from the 
	University of Auckland, New Zealand. He is currently 
	a Senior Lecturer at the University of Auckland. 
	His research interests include embedded systems, 
	robotics, artificial neural networks, control systems, 
	and hardware implementation of real-time systems for 
	control.
\end{IEEEbiography}





\end{document}